\documentclass[12pt]{article}
\usepackage{newinutile}
\usepackage{amsmath,amssymb,amsthm,amsfonts}
\usepackage{epsfig,graphics,color,calc,graphicx}
\usepackage{subfigure}
\graphicspath{{Figures/}}
\usepackage{comment}
\usepackage{pgf,tikz}
\usepackage{mathrsfs}
\usepackage{color}
\usetikzlibrary{arrows}
\usepackage{cancel}
\usepackage{mathrsfs}
\usepackage{hyperref}
\hypersetup{colorlinks=true,linkcolor=blue}
\usepackage{placeins}

\usepackage{cite}
\usepackage{mathrsfs}

\def\theequation{\arabic{section}.\arabic{equation}}

\newcommand{\rexit}{\textup{RE}}

\newlength{\pecettawidth}
\begin{document}
\title{Residence time in presence of moving defects and obstacles}

\author{E.N.M.\ Cirillo}
\email{emilio.cirillo@uniroma.it}
\affiliation{Dipartimento di Scienze di Base e Applicate per l'Ingegneria,\\
             Sapienza Universit\`a di Roma, \\
             via A.\ Scarpa 16, 00161, Roma, Italy.}

\author{Matteo Colangeli}
\email{matteo.colangeli1@univaq.it}
\affiliation{Dipartimento di Ingegneria e Scienze dell'Informazione e
Matematica,
Universit\`a degli Studi dell'Aquila, via Vetoio, 67100 L'Aquila, Italy.}

\author{Antonio Di Francesco}
\email{antonio.difrancesco5@graduate.univaq.it}
\affiliation{Dipartimento di Ingegneria e Scienze dell'Informazione e
Matematica,
Universit\`a degli Studi dell'Aquila, via Vetoio, 67100 L'Aquila, Italy.}


\begin{abstract}
We discuss the properties of the residence time in 
presence of moving defects or obstacles for a particle 
performing a one dimensional random walk. More precisely,
for a particle conditioned to exit through the right endpoint, 
we measure the typical time needed to cross
the entire lattice in presence of defects. 
We find explicit formulae for the residence time and discuss
several models of moving obstacles. The presence of a stochastic
updating rule for the motion of the obstacle smoothens the 
local residence time profiles found in the case of a static obstacle.
We finally discuss connections with applicative problems,
such as the pedestrian motion in presence of queues and the residence time
of water flows in runoff ponds.      
\end{abstract}




\maketitle

\section{Introduction}
\label{s:intro}
\par\noindent
The characterization of transport regimes in presence of obstacles and irregular patterns is a classical problem of fluid dynamics \cite{Lane95,Will96}, which is also relevant to a number of biological \cite{Chep19} and engineering  \cite{Creed17,Kirby13} applications. For instance, a careful design of the shape and the location of the obstacles in microfluidic channels was observed to enhance the mixing of fluid flows \cite{Wang02}.
Moreover, crowding effects, generated by obstacles of different size and shape, turn out to significantly affect the transport properties by even altering the sign of the fluxes, as recently reported, e.g., in active matter numerical experiments \cite{Simpson16,Borba20}.

From the perspective of statistical mechanics, the subject is particularly rich and may be tackled from different routes.
One first important question, for instance, concerns the study of anomalous diffusions resulting from the presence of obstacles \cite{Saxton94,Berry14}. The onset of anomalous transport stemming from the motion of intruders in a host matrix of slowly moving particles was studied in \cite{Sciort16}.
The inclusion of flower shape obstacles in billiards was considered in \cite{Klages02} in order to highlight the fractal structure of the diffusion coefficient as a function of control parameters. 
Another interesting research line concerns the derivation of the hydrodynamic limit of interacting particle systems in an inhomogeneous environment. In \cite{ACCG19} it was shown that a detailed characterization of the inhomogeneities at the microscopic level is necessary to determine the structure of the macroscopic diffusion laws in the hydrodynamic limit. Furthermore, in the framework of Zero Range Processes, the inclusion of local defects proved to be a mechanism able to give rise to condensation phenomena in the hydrodynamic limit \cite{CC16}.

The investigation of inhomogeneous random walks continues to be a core topic in statistical mechanics, see e.g. \cite{Mesn} for a general review of the subject.

In this work we study the residence time in a random walk on a one-dimensional lattice in presence of local inhomogeneities. The walker is conditioned to exit through the right endpoint of the lane. 
Our approach permits explicit computations that nicely complement the classical literature based on the gambler's ruin problem. 
We consider both symmetric and asymmetric random walks, and discuss the residence time in presence of defects, represented by lattice sites on which the hopping probabilities of the walker are perturbed by a \textit{bias}. In particular, analytical formulae for the residence time are obtained for random walks with static defects as well as in presence of defects following a stochastic updating rule. 

By exploiting the theory of absorbing Markov chains, we succeed, to a certain general extent, to express the residence time of the conditioned random walk in terms of quantities evaluated in the non conditioned walk: the mean number of visits on the generic site of the lattice and the right exit probability. Thus, we dwell on the derivation of analytic formulae for these two quantities and use them to write explicit formulae for the residence time in presence of static defects. The prominent effect of the inclusion of a defect obeying a stochastic dynamics is the smoothing of the local residence time profiles. Our analysis, supported by numerical simulations, also shows that the presence of a stochastic defect may induce a decrease of the residence time. Moreover, for certain distributions of the defect, the residence time turns out to be invariant with respect to the sign of the bias acting on the defect.

Our results are also amenable to a variety of applications. We discuss, in particular, the use of our formalism in the modelling of pedestrians forming queues and also outline useful links with hydrogeological problems. In both cases, we show that the results found in those contexts by means of sophisticated models
can be recovered to some extent in the framework of our one dimensional random walk model through a suitable choice of the parameters.

In addition, we show in the Appendix how our results on the stationary local residence time profile connect with the stationary fugacity profiles of arbitrary Zero Range Processes subject to specific  injection rates at the boundaries.

Our work is organized as follows. In Sec. \ref{s:mod}, we introduce the theoretical set-up for the computation of the residence time, based on the \textit{fundamental matrix}, and define the main quantities of interest. In Sec. \ref{s:explicit} we apply our formalism to both symmetric and asymmetric random walks and obtain explicit formulae in presence of static defects. In Sec. \ref{s:move} we study the residence time problem in presence of moving defects which follow a stochastic updating rule. The discussion of relevant applications and our conclusions are deferred to Sec. \ref{s:con}, while the analogy with Zero Range Processes is outlined in the Appendix \ref{a:occup}.

\section{Model and quantities of interest}
\label{s:mod}
\par\noindent
The problem discussed above is approached in this paper via
a one--dimensional model based on a simple random walk.
At each side of the lattice an absorbing site is present.
We shall first consider a completely general simple random walk and 
then we will specialize our discussion to particular 
models describing the effect of static or moving defects.

We thus consider a one--dimensional simple random walk 
on $\{0,1,\dots,L\}$, called 
\emph{lane}. Each element of the lane will be called \emph{site}.
The sites $0$ and $L$ are absorbing,
so that when the particle reaches one of them the walk is stopped.
They will be called, respectively, the \emph{left} and \emph{right exit} of 
the lane. The sites $1,\dots,L-1$ are called \emph{transient}.

Time $t=0,1,2,\dots$ is discrete
and 
at each time the walker
jumps from a transient site $i$ 
to site $j$ with probability $p_{ij}$.
We assume $p_{ij}=0$ if $|i-j|=0$ or $|i-j|\ge2$ and $p_{ij}>0$ 
otherwise. Moreover, we shall use the notation 
$p_{i i+1}=p_i$ 
and 
$p_{i i-1}=q_i$, 
for any $i=1,\dots,L-1$. We will obviously have that $p_i+q_i=1$.
In other words, at each time the walker jumps from the transient 
site $i$ to its 
left neighbor $i-1$ with probability $q_i$ or to its right neighbor $i+1$
with probability $p_i=1-q_i$.

We denote by $X(t)$ the position of the walker at time $t$ and
by $\mathbb{P}_i$ and $\mathbb{E}_i$ the probability associated to the
process and the related expected value operator
for the walk started at $X(0)=i$, with $i=1,\dots,L-1$. 

We denote by $Q$ the 
$(L-1)\times(L-1)$ 
square matrix in which we collect the transition 
probabilities between transient sites, namely, 
$Q_{ij}=p_{ij}$ for $i,j=1,\dots L-1$.
The matrix 
\begin{equation}
\label{def010}
\mathbb{I}-Q
=
	 \begin{bmatrix}
	 1 & -p_1 & 0 & \cdots \\
	 -q_2 & 1 & -p_2 & 0 & \cdots \\
	 0 & -q_3 & 1 & -p_3 & 0 & \cdots\\
	 \vdots & & & \ddots\\
	 & & \cdots & 0 & -q_{L-2} & 1 & -p_{L-2}\\
	 & & & \cdots & 0 & -q_{L-1} & 1 
	 \end{bmatrix}
\end{equation}
is invertible, see, for instance, \cite[Theorem~3.2.1]{KS1960}, and 
its inverse $N=(\mathbb{I}-Q)^{-1}$ is called the \emph{fundamental matrix}.

\subsection{Residence time}
\label{s:res}
\par\noindent
The residence time is defined by starting
the walk at site $1$ and computing
the typical time that the particle takes to reach the site $L$ provided
the walker reaches $L$ before $0$.
More precisely, we denote by $\tau=\inf\{t>0:\,X(t)\in\{0,L\}\}$
the \emph{duration of the walk}
and 
define the 
\emph{residence time},
as 
\begin{equation}
\label{def000}
\Gamma=\mathbb{E}_1(\tau|\rexit),
\end{equation}
where we conditioned to the event $\rexit$ meaning that the
particle exits the lane through the right exit in $L$.

We shall compute the residence time following ideas differing
from the approaches already developed in \cite{CC19,CCSpre2018,CCVpre2020}. 
In particular, we shall exploit several properties of the fundamental matrix 
of absorbing Markov chains \cite{KS1960}
that will allow us in Section~\ref{s:explicit}
to derive remarkable explicit formulae. 

Given a site $i$, the 
\emph{number of visits} 
$n_i=|\lbrace t>0 : X(t)=i \rbrace |$
to $i$ counts the number of times that the walker visits 
the site $i$. 
Its mean value for the walk started at site $j$ can be computed 
using the fundamental matrix, indeed one has that 
$\mathbb{E}_j[n_i]=N_{ji}$, see \cite[Theorem~3.2.4]{KS1960}.
The mean number of visits of the walker started at $1$ 
conditioned to the event that it exits the lane in $L$, namely, 
$\mathbb{E}_1[n_i|\rexit]$, is called 
\emph{local residence time} at $i$. 
Since, for the walker started at $1$, $\tau=\sum_{i=1}^{L-1}n_i$, 
the residence time can be written as sum 
of the local residence times, namely,
\begin{equation}
\label{def020}
\Gamma
=
\sum_{i=1}^{L-1}
\mathbb{E}_1[n_i|\rexit].
\end{equation}

Using the standard theory of absorbing Markov chains it is possible 
to get rid of the conditioning, as it is can be proven 
that
\begin{equation}
\label{def030}
\mathbb{E}_1[n_i|\rexit]
=
\frac{\mathbb{P}_i[\rexit]}{\mathbb{P}_1[\rexit]}
\;
\mathbb{E}_1[n_i]
.
\end{equation}
Indeed, 
we consider the random walk conditioned to exit the lane in $L$, 
which is a Markov chain with the single absorbing state $L$.
The fundamental matrix of such a random walk, 
see \cite[Chapter~III, page~65]{KS1960}, 
is given by $D^{-1} N D$, where $D$ is the $(L-1)\times(L-1)$ diagonal 
matrix whose elements are $\mathbb{P}_i[\rexit]$ for 
$i=1,\dots,L-1$.
The inverse $D^{-1}$ is diagonal and its elements are 
$1/\mathbb{P}_i[\rexit]$. Thus, 
for $i$ fixed 
\begin{displaymath}
(D^{-1}ND)_{1i}
=
D^{-1}_{1s}N_{sr}D_{ri}
=
\frac{1}{\mathbb{P}_1[\rexit]}
N_{1r}D_{ri}
=
\frac{1}{\mathbb{P}_1[\rexit]}
N_{1i}
\mathbb{P}_i[\rexit]
,
\end{displaymath}
which proves \eqref{def030}.

Using the equality \eqref{def030} the residence time computation 
can be reduced to the computation of properties of the 
non conditioned random walk, indeed
\begin{equation}
\label{def040}
\Gamma
=
\frac{1}{\mathbb{P}_1[\rexit]}
\sum_{i=1}^{L-1}
\mathbb{P}_i[\rexit]
\mathbb{E}_1[n_i]
.
\end{equation}

\subsection{Mean number of visits}
\label{s:vis}
\par\noindent
Since, as mentioned above, $\mathbb{E}_1[n_i]=N_{1i}$, in order to 
compute the mean number of visits of the walker at the generic transient 
site $i$, we have to compute the first raw of the fundamental matrix, 
which, we recall, is
the inverse of the tridiagonal matrix $\mathbb{I}-Q$ 
given in \eqref{def010}. This can be done in several different ways, 
here we follow the approach proposed in \cite{U1994}, which 
will show to be very powerful to deduce the explicit formulae 
that we shall derive in Section~\ref{s:explicit}. Moreover, 
this approach will also allow an elegant derivation of the 
analogy that we shall discuss in the Appendix~\ref{a:occup}.

First we note that from \cite[equation~(1.2)]{U1994}
the determinant of the matrix $\mathbb{I}-Q$ is equal 
to the last term $\theta_{L-1}$ of the sequence $\theta_i$ defined 
by the following recursive equations
\begin{equation}
\label{def045}
\left\{
\begin{array}{l}
\theta_i=\theta_{i-1}-q_ip_{i-1}\theta_{i-2}
\;\;\;\;
i=1,2,\dots,L-1\\
\theta_{-1}=0,\;\theta_{0}=1.
\end{array}
\right.
\end{equation}
We borrow from \cite{U1994} also the definition of the sequence 
$\phi_i$ given through the recursive equations 
\begin{equation}
\label{def060}
\left\{
\begin{array}{l}
\phi_i=\phi_{i+1}-q_{i+1}p_i\phi_{i+2}
\;\;\;\;
i=L-1,L-2,\dots,2,1\\
\phi_{L}=1,\;\phi_{L+1}=0.
\end{array}
\right.
\end{equation}
The sequences $\theta_i$ and $\phi_i$ are not independent, indeed, 
\cite[Lemma~2]{U1994} states that 
\begin{equation}
\label{def070}
\theta_i\phi_{i+1}-q_{i+1}p_i\theta_{i-1}\phi_{i+2}=\theta_{L-1}
\;\;\;\;
i=L-1,L-2,\dots,2,1.
\end{equation}
In particular, since from \eqref{def045} $\theta_1=1$, 
comparing \eqref{def070} and \eqref{def060} both for $i=1$, 
we get 
\begin{equation}
\label{def080}
\phi_1=\theta_{L-1}. 
\end{equation}

Now, 
by using \cite[Lemma~4 and Theorem~2]{U1994} and exploiting 
the equality 
\eqref{def080}, we have that 
\begin{equation}
\label{def050}
N_{1i}
=
\frac{1}{\phi_1}
\phi_{i+1}
\prod_{k=1}^{i-1}p_k
\;\;\;\;\;\;
i=1,\dots,L-1
.
\end{equation}

Equations \eqref{def050} and \eqref{def060} can be combined to derive 
a set of recursive equations for $N_{1j}$. Indeed, 
using \eqref{def060} for $i=1$ 
and \eqref{def050} for $i=1,2$,
we have 
\begin{equation}
\label{def090}
1
=
\frac{1}{\phi_1}\phi_2-\frac{1}{\phi_1}q_2p_1\phi_3
=
N_{11}
-
q_2 N_{12}.
\end{equation}
Using \eqref{def060} for $i=L-1$,
and \eqref{def050} for $i=L-1,L-2$,
we get
\begin{equation}
\label{def100}
\phi_{L-1}
=
1
\Rightarrow
\frac{1}{\phi_1}
\phi_{L-1}
\prod_{k=1}^{L-3}p_k
=
\frac{1}{\phi_1}
\prod_{k=1}^{L-3}p_k
\Rightarrow
N_{1L-2}
=
\frac{1}{p_{L-2}}
N_{1L-1}.
\end{equation}
Now, we consider \eqref{def060} for $i=L-2,\dots,2$
and multiply it for a suitable coefficient
to obtain 
\begin{displaymath}
\frac{1}{\phi_1}
\phi_{i}
\prod_{k=1}^{i-1}p_k
=
\frac{1}{\phi_1}
\phi_{i+1}
\prod_{k=1}^{i-1}p_k
-
q_{i+1}p_i
\frac{1}{\phi_1}
\phi_{i+2}
\prod_{k=1}^{i-1}p_k
.
\end{displaymath}
Exploiting \eqref{def050} for $i=L-1,L-2,\dots,1$,
we get
\begin{equation}
\label{def110}
N_{1i-1}p_{i-1}
=
N_{1i}
-
q_{i+1}
N_{1i+1}
.
\end{equation}

The derivation of equations \eqref{def110} is the key point of our computation: 
the recursive equations for $\theta_i$ and $\phi_i$, thanks to the introduction of
$N_{1i}$, have been recast in a form which can be easily solved once
it is rewritten in the form of a current conservation law.
Indeed, by exploiting the fact 
that $p_i+q_i=1$, we rewrite the whole set of equations 
\eqref{def090}--\eqref{def110} as
\begin{equation}
\label{def120}
1-q_1N_{11}
=
p_1N_{11}-q_2N_{12}
=
\cdots
=
p_{L-2}N_{1L-2}-q_{L-1}N_{1L-1}
=
p_{L-1}N_{1L-1}
.
\end{equation}
Thus, the recursive equations \eqref{def110}, together with the boundary 
equations \eqref{def090} and \eqref{def100}, allow to find the 
mean visit number profile $N_{1i}$ for $i=1,\dots,L-1$. 

We denote by $c$ the conserved quantity defined by equations \eqref{def120}
and solve by induction the first $L-1$ equations 
\begin{displaymath}
1-q_1N_{11}
=
p_1N_{11}-q_2N_{12}
=
\cdots
=
p_{L-2}N_{1L-2}-q_{L-1}N_{1L-1}
=c
\end{displaymath}
obtaining\footnote{\label{f:convention}We remark that here, and in the 
following, we shall 
always adopt the convention that the sum and the product symbols 
mean, respectively, $0$ and $1$ when the index corresponding to the 
first element is greater than the one corresponding to the last one.}
\begin{equation}
\label{def130}
N_{1i}
=
\frac{\prod_{k=1}^{i-1}p_k}{\prod_{k=1}^iq_k}
-
c
\sum_{s=0}^{i-1}
\frac{\prod_{k=s+1}^{i-1}p_k}{\prod_{k=s+1}^iq_k}
\;\;\;\;\;
i=1,\dots,L-1.
\end{equation}
The last equation $p_{L-1}N_{1L-1}=c$ of the set of equations 
\eqref{def120} can be 
then used to set the value of the constant $c$, indeed, 
from 
\begin{displaymath}
p_{L-1}
\bigg[
\frac{\prod_{k=1}^{L-2}p_k}{\prod_{k=1}^{L-1}q_k}
-
c
\sum_{s=0}^{L-2}
\frac{\prod_{k=s+1}^{L-2}p_k}{\prod_{k=s+1}^{L-1}q_k}
\bigg]
=c,
\end{displaymath}
we get 
\begin{equation}
\label{def140}
c
=
\frac{\prod_{k=1}^{L-2}p_k}{\prod_{k=1}^{L-1}q_k}
\bigg[
\frac{1}{p_{L-1}}
+
\sum_{s=0}^{L-2}
\frac{\prod_{k=s+1}^{L-2}p_k}{\prod_{k=s+1}^{L-1}q_k}
\bigg]
^{-1}
.
\end{equation}

We remark that, summing $N_{1i}$ for $i=1,\dots,L-1$, one can use 
\eqref{def130} to compute the total length of the walk that, in the 
gambler's ruin language, is the duration of the game. 
It is a straightforward exercise to show that for the 
symmetric walk, namely, $p_i=q_i=1/2$ for $i=1,\dots,L-1$,
one finds $c=1/L$ and $N_{1i}=2-2i/L$, so that the duration of the walk 
is $2\sum_{i=1}^{L-1}(1-i/L)=L-1$, see  
\cite[equation~(3.5) in Chapter~XIV]{Fel}.
The computation is more involved in the homogeneous driven case, 
i.e., $p_i=p$ and $q_i=q$ for $i=1,\dots,L-1$ with $p\neq q$. 
In such a case one finds 
$c=(p^{L-2}/q^{L-1})[1/p+(p^{L-2}/q^{L-1})((q/p)^{L-1}-1)/(q/p-1)]^{-1}$
and $N_{1i}=(p^{i-1}/q^i)[1-c((q/p)^i-1)/(q/p-1)]$, 
and summing from $1$ to $L-1$ we find the length of the walk
$1/(q-p)-(L/(q-p))(1-q/p)/(1-(q/p)^L)$,
see \cite[equation~(3.4) in Chapter~XIV]{Fel}.

\subsection{Right exit probability}
\label{s:rep}
\par\noindent
The probability of absorption by one particular absorbing state is a classical 
topic both in the gambler's ruin problem \cite{Fel,CC19} 
and in the absorbing Markov chains literature. 
Adopting the gambler's ruin point of view, we let 
$t_i=1-\mathbb{P}_i[\rexit]$ be the probability that the walker started 
at $i$ exits the lane at $0$, namely, the probability that the 
gambler with initial fortune $i$ is eventually ruined, and 
we note that 
\begin{equation}
\label{def150}
\left\{
\begin{array}{l}
t_i=q_it_{i-1}+p_it_{i+1}
\;\;\;\;
i=2,\dots, L-2\\
t_1=q_1+p_1t_2\\
t_{L-1}=q_{L-1}t_{L-2}.\\
\end{array}
\right.
\end{equation}

We just mention that the same set of equations can be found 
taking the absorbing Markov chain point of view, indeed
we let $R$ be the $(L-1)\times 1$ 
column vector collecting the probability to jump from any transient site 
to $L$, i.e., $R_{i1}=0$ if $i=1,\dots,L-2$ and $R_{L-1 1}=p_{L-1}$. 
Moreover, we consider the $(L-1)\times 1$ column vector $B$ such that 
$B_{i1}=\mathbb{P}_i[\rexit]$ is the probability that the walker started 
at site $i$ ends its walk in $L$. From \cite[Theorem~3.3.7]{KS1960}
we have that $B=NR$, so that 
\begin{equation}
\label{def160}
B_{i1}
=
N_{ij}R_{j1}
=
N_{iL-1}p_{L-1}.
\end{equation}
Thus, the computation of the right exit probability is reduced 
to that of the last column of the fundamental matrix. A computation 
similar to the one performed in Section~\ref{s:vis} yields the recursive 
equations for $B_{i1}$ analogous to \eqref{def150}.

Now, we come back to the study of 
the recursive equations \eqref{def150}, which 
can be rewritten in the more compact form
\begin{equation}
\label{def170}
\left\{
\begin{array}{l}
t_i=q_it_{i-1}+p_it_{i+1}
\;\;\;\;
i=1,\dots, L-1\\
t_0=1, \;\; t_{L}=0.\\
\end{array}
\right.
\end{equation}
To solve \eqref{def170} we first note that the sequences 
$s_i=1$ and $r_i=\sum_{k=1}^i\prod_{r=k}^{L-1}p_r/q_r$ are 
solutions of the recursive equations, but do not satisfy the 
boundary conditions (recall the footnote~\ref{f:convention}).
The statement is trivial for $s_i$, while it requires 
some effort for $r_i$:
\begin{displaymath}
q_ir_{i-1}+p_ir_{i+1}
=
q_i
\sum_{k=1}^{i-1}\prod_{r=k}^{L-1}\frac{p_r}{q_r}
+
p_i
\sum_{k=1}^{i+1}\prod_{r=k}^{L-1}\frac{p_r}{q_r}
=
q_i
\sum_{k=1}^{i-1}\prod_{r=k}^{L-1}\frac{p_r}{q_r}
+
p_i
\sum_{k=1}^{i}\prod_{r=k}^{L-1}\frac{p_r}{q_r}
+
p_i
\prod_{r=i+1}^{L-1}\frac{p_r}{q_r}
\end{displaymath}
and thus
\begin{displaymath}
q_ir_{i-1}+p_ir_{i+1}
=
q_i
\sum_{k=1}^{i-1}\prod_{r=k}^{L-1}\frac{p_r}{q_r}
+
p_i
\sum_{k=1}^{i}\prod_{r=k}^{L-1}\frac{p_r}{q_r}
+
q_i
\prod_{r=i}^{L-1}\frac{p_k}{q_k}
=
q_i
\sum_{k=1}^{i}\prod_{r=k}^{L-1}\frac{p_r}{q_r}
+
p_i
\sum_{k=1}^{i}\prod_{r=k}^{L-1}\frac{p_r}{q_r}
,
\end{displaymath}
which is equal to $r_i$ as $q_i+p_i=1$. 
Thus, exploiting the recursive linearity of the equation \eqref{def170}, 
we can look for a solution satisfying the boundary condition 
in the form $t_i=a s_i+b r_i$, where $a,b$ are real constants. 
It is not difficult to verify that $a=1$ and 
$b=-[1+\sum_{k=1}^{L-1}\prod_{r=k}^{L-1}p_r/q_r]^{-1}$ do the job, 
so that the sought for solution of the system \eqref{def170} is
\begin{equation}
\label{def180}
t_i
=
1
-
\frac{1}{1+\sum_{k=1}^{L-1}\prod_{r=k}^{L-1}\frac{p_r}{q_r}}
\sum_{k=1}^{i}\prod_{r=k}^{L-1}\frac{p_r}{q_r}
.
\end{equation}
It is a straightforward exercise to check that in the homogeneous 
case, in which the jumping probabilities are the same at each site, 
the classical results of the gambler's ruin problem (see, e.g., 
\cite[equations~(2.4) and (2.5) in Chapter~XIV]{Fel} or 
\cite[Section~3.1]{CC19}) are recovered.

Finally, recalling that $t_i$ is the probability that the 
walker started at $i$ exits the lane at $0$, we have that 
the right exit probability is given by 
\begin{equation}
\label{def190}
\mathbb{P}_i[\rexit]
=
\frac{1}{1+\sum_{k=1}^{L-1}\prod_{r=k}^{L-1}\frac{p_r}{q_r}}
\sum_{k=1}^{i}\prod_{r=k}^{L-1}\frac{p_r}{q_r}
.
\end{equation}

\section{Explicit expression of the residence time in presence 
of a static defect}
\label{s:explicit}
\par\noindent
In this section we suppose that 
all the sites $1,\dots,L-1$ share the same behavior, i.e., they are 
\emph{regular}, save for one
site called \emph{defect}
\cite{CC19,CCSpre2018}.
The defect site is the site $2\le d\le L-2$.
More precisely, 
we assume that $p_i=p$ and $q_i=q=1-p$ for all 
$i=1,\dots,L-1$ such that $i\neq d$ and 
$p_d=\bar{p}$ and $q_d=\bar{q}=1-\bar{p}$. 
We shall also write 
$\bar{p}=p+\epsilon$ and $\bar{q}=q-\epsilon$, 
with $\epsilon\in(-p,q)$ called \textit{bias}.
We will call \emph{symmetric} the case in which $p=q$ 
and \emph{driven} that in which $p\neq q$. In the driven
case $p-q$ is called \emph{drift}. 

Note that, if $\epsilon=0$ the defect is not present and 
the classical gambler ruin problem 
is recovered.
The case $\epsilon=-p$ cannot be considered, because it would mean that 
the walker is reflected to the left 
with probability one by the defect and, hence, 
it would never reach the right--end exit in $L$. 
The case $\epsilon=p$ is meaningful, but rather extreme, 
indeed, in such a case, the defect pushes the walker to the right 
with probability one and this means that, once the walker started at 
$1$ has overcome the defect, it will never come back to the 
part of the lane on the left of the defect. 
In any case this case is not covered by our study. 

\subsection{The symmetric case}
\label{s:sym}
\par\noindent
We start by computing $c$ given in \eqref{def140}.
In the first factor all the terms cancel but the one associated with the 
defect and the last term $q_{L-1}$ at the denominator. The sum in the 
second 
factor behaves similarly if $s$ ranges from $0$ to $d-1$.
On the other hand, for the remaining $L-2-(d-1)$ terms corresponding 
to $s$ ranging from $d$ to $L-2$, the ratio results equal to 
$1/q_{L-1}$. Thus, we have 
\begin{displaymath}
c
=
\frac{2(1/2+\epsilon)}{(1/2-\epsilon)}
\bigg[
      2
      +\frac{2(1/2+\epsilon)}{(1/2-\epsilon)}d
      +2(L-2-d+1)
\bigg]^{-1}
,
\end{displaymath}
which yields 
\begin{equation}
\label{exp000}
c=\frac{1+2\epsilon}{L(1-2\epsilon)+4\epsilon d}.
\end{equation}

To compute the mean number of visits profile \eqref{def130} we use similar 
arguments, but we have to distinguish three cases. We get
\begin{equation}
\label{exp010}
N_{1i}
=
\left\{
\begin{array}{ll}
2
-
2ci
&
\;\;\;\;\;
i=1,\dots,d-1
\\
{\displaystyle
\frac{2}{1-2\epsilon}
( 1 - cd )
\vphantom{\bigg\{_\big\}}
}
&
\;\;\;\;\;
i=d
\\
{\displaystyle
\frac{2(1+2\epsilon)}{1-2\epsilon}
-c\frac{8\epsilon d}{1-2\epsilon}
-
2ci
\vphantom{\bigg\{_\big\}}
}
&
\;\;\;\;\;
i=d+1,\dots,L-1.
\\
\end{array}
\right.
\end{equation}

The following step is the computation of the right exit probability. 
From \eqref{def190}, distinguishing two different cases, we have 
\begin{equation}
\label{exp040}
\mathbb{P}_i[\rexit]
=
\left\{
\begin{array}{ll}
{\displaystyle
\frac{(1+2\epsilon)i}{L(1-2\epsilon)+4\epsilon d} 
\vphantom{\bigg\{_\big\}}
}
&
\;\;\;
i\le d\\
{\displaystyle
\frac{(1-2\epsilon)i}{L(1-2\epsilon)+4\epsilon d} 
+
\frac{4\epsilon d}{L(1-2\epsilon)+4\epsilon d} 
}
&
\;\;\;
i> d.\\
\end{array}
\right.
\end{equation}

In view of computing the residence time, 
we now use \eqref{def030} to give an explicit expression of the 
local residence time profile:
\begin{equation}
\label{exp050}
\mathbb{E}_i[n_i|\rexit]
=
\left\{
\begin{array}{ll}
2
i
-
2ci^2
&
\;\;\;\;\;
i=1,\dots,d-1
\\
{\displaystyle
\frac{2d}{1-2\epsilon}
( 1 - cd )
\vphantom{\bigg\{_\big\}}
}
&
\;\;\;\;\;
i=d
\\
{\displaystyle
\frac{(1-2\epsilon)i+4\epsilon d}{1+2\epsilon}
\bigg[
\frac{2(1+2\epsilon)}{1-2\epsilon}
-c\frac{8\epsilon d}{1-2\epsilon}
-
2ci
\bigg]
\vphantom{\bigg\{_\big\}}
}
&
\;\;\;\;\;
i=d+1,\dots,L-1.
\\
\end{array}
\right.
\end{equation}
Finally, \eqref{def020} is used to compute the residence time
\begin{equation}
\label{exp060}
\begin{array}{rcl}
\Gamma
&\!\!=&\!\!
{\displaystyle
(1-c)(d-1)d
+
\frac{2d(1-cd)}{1-2\varepsilon}
}
\\
&&\!\!
{\displaystyle
-\frac{L-d-1}{3(1-4\epsilon^2)}
(-3d+cd+2cd^2-3L-cL+2cdL+2cL^2-24d\epsilon-4cd\epsilon
}
\\
&&\!\!
{\displaystyle
\phantom{-\frac{L-d-1}{3(1-4\epsilon^2)}(}
+16cd^2\epsilon+4cL\epsilon+16cdL\epsilon-8cL^2\epsilon-36d\epsilon^2
+4cd\epsilon^2+56cd^2\epsilon^2
}
\\
&&\!\!
{\displaystyle
\phantom{-\frac{L-d-1}{3(1-4\epsilon^2)}(}
+12L\epsilon^2-4cL\epsilon^2-40cdL\epsilon^2+8cL^2\epsilon^2)
}
\\
&\!\!=&\!\!
{\displaystyle
\frac{L^3(1-2\epsilon)+12 d \epsilon L^2
      -L(1+24d^2\epsilon-2\epsilon)+16d^3\epsilon-4d\epsilon}
     {3[L(1-2\epsilon)+4\epsilon d]}
.
}
\end{array}
\end{equation}
The leading term for $L$ large, uniformly in the choice of the 
other parameters $d$ and $\epsilon$, is $L^2/3$. 
Thus, the presence 
of the defect does not affect for large $L$ 
the diffusive character of the walk.

Finally, 
it is worth noting that equation \eqref{exp060}, in the case $\epsilon=0$, 
reduces to the residence time for the symmetric model in absence of defect, 
namely, for the gambler's ruin problem. Indeed we get
$\Gamma=(L^2-1)/3$ as also shown in \cite[Section~IV.A]{CCSpre2018}.

\begin{figure}[!h]
	\centering
	\includegraphics[width=0.475\textwidth]{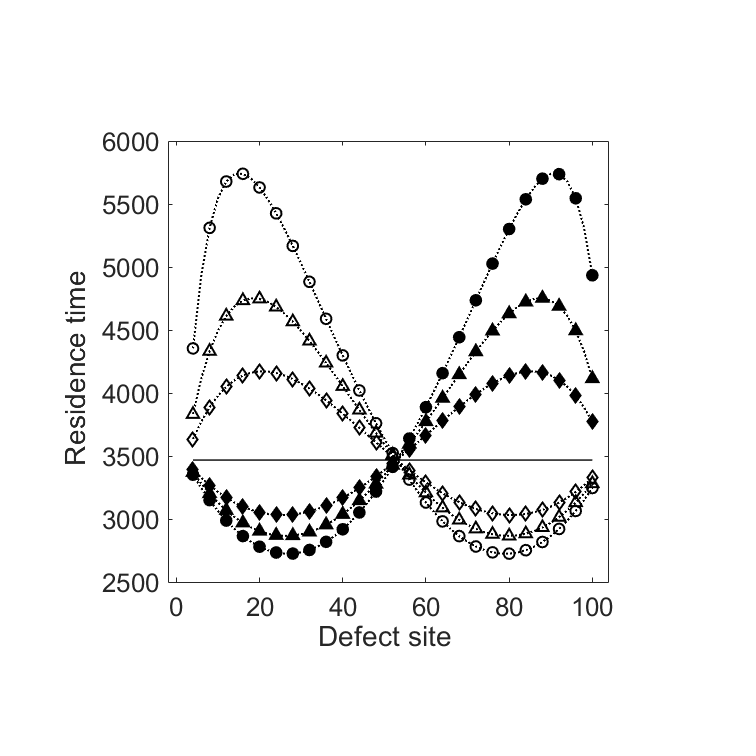}
	\caption{Residence time as a function of $d$ for a symmetric random walk (i.e., $p=q=0.5$) for $L=101$. In particular, plain circles, triangles and diamonds refer to the defect intensities $\epsilon=-0.4,\: -0.3,\: -0.2$ respectively, while same symbols but empty refer to $\epsilon=0.4,\: 0.3,\: 0.2$. The solid line corresponds to the residence time of the system without any defect, namely $\Gamma=3467.7$ time steps.}
	\label{fig4.1:NullModel_Gamma_NONdriven}
\end{figure}

Note that the model presented and analyzed so far is analogous to the one studied by Ciallella et al. in \cite{CC19}. Fig. \ref{fig4.1:NullModel_Gamma_NONdriven} shows the dependence of the residence time $\Gamma$ on the defect site position $d$ in a symmetric random walk, computed through the use of formula \eqref{exp060}.

In particular, Fig. \ref{fig4.1:NullModel_Gamma_NONdriven} shows that the behavior of the residence time as a function of the defect site $d$ is  analogous to that already observed in \cite{CC19}. In fact, the residence time is maximized or minimized by the position of the defect, depending on the sign of $\epsilon$, and it is not affected by the defect intensity when $d$ coincides with the central site of the lattice. 

\subsection{The driven case}
\label{s:dri}
\par\noindent
We start by computing $c$ given in \eqref{def140}.
In the first factor all the terms contribute with the ratio $p/q$ 
but the one associated with the 
defect, which is $\bar{p}/\bar{q}$, and the last term 
$q_{L-1}$ at the denominator. 
The sum in the 
second factor behaves similarly if $s$ ranges from $0$ to $d-1$.
On the other hand, for the remaining $L-2-(d-1)$ terms corresponding 
to $s$ ranging from $d$ to $L-2$, the ratio results equal to 
$(p/q)^{L-2-s} (1/q)$. 

In order to make the following formulae more compact and legible,
we will use the shorthand notation 
\begin{displaymath}
A=\frac{q}{p}
\;\;\;\textup{ and }
\;\;\;
\bar{A}=\frac{\bar{q}}{\bar{p}}
.
\end{displaymath}
With this convention, 
we have 
\begin{displaymath}
c
=
A^{3-L}
\frac{1}{\bar{A}}
\frac{1}{q}
\bigg[
      \frac{1}{p}
      +
      \sum_{s=0}^{d-1}
        A^{s+3-L}
        \frac{1}{\bar{A}}
        \frac{1}{q}
      +
      \sum_{s=d}^{L-2}
        A^{s+2-L}
        \frac{1}{q}
\bigg]^{-1}
,
\end{displaymath}
which, after some algebra, yields 
\begin{equation}
\label{exp070}
c
=
A^{3-L}
\frac{1}{\bar{A}}
\frac{1}{q}
\bigg[
      \frac{1}{p}
      +
      A^{3-L}
      \frac{1}{\bar{A}}
      \frac{1}{q}
      \frac{A^d-1}{A-1}
      +
      A^{2-L}
      \frac{1}{q}
      \frac{A^{L-1}-A^d}{A-1}
\bigg]^{-1}
.
\end{equation}

To compute the mean number of visits profile \eqref{def130} we use similar 
arguments, but we have to distinguish three cases. We get 
\begin{equation}
\label{exp080}
N_{1i}
=
\left\{
\begin{array}{ll}
{\displaystyle
A^{1-i}\frac{1}{q}\Big[1-c\frac{A^i-1}{A-1}\Big]
\vphantom{\bigg\{_\big\}}
}
&
\;\;\;\;\;
i=1,\dots,d-1
\\
{\displaystyle
A^{1-d}\frac{1}{\bar{q}}\Big[1-c\frac{A^d-1}{A-1}\Big]
\vphantom{\bigg\{_\big\}}
}
&
\;\;\;\;\;
i=d
\\
{\displaystyle
A^{2-i}\frac{1}{\bar{A}}\frac{1}{q}
-c
\Big[
     A^{2-i}\frac{1}{\bar{A}}\frac{1}{q}
     \frac{A^d-1}{A-1}
     +
     A^{1-i}\frac{1}{q}
     \frac{A^i-A^d}{A-1}
\Big]
\vphantom{\bigg\{_\big\}}
}
&
\;\;\;\;\;
i=d+1,\dots,L-1.
\\
\end{array}
\right.
\end{equation}
Note that, due to the presence of the term $1/\bar{q}$ in the front 
factor, the second case is not simply a particularization of 
the first one. 

The following step is the computation of the right exit probability. 
Although not strictly necessary to derive the expression of the 
residence time, we first compute the front factor appearing in  
\eqref{def190}:
\begin{equation}
\label{exp090}
Z
=
1+\sum_{k=1}^{L-1}\prod_{r=k}^{L-1}\frac{p_r}{q_r}
=
1
+
\frac{1}{\bar{A} A^L}
\frac{A^{d+1}-A}{A-1}
+
\frac{1}{A^L}
\frac{A^L-A^{d+1}}{A-1}.
\end{equation}
Distinguishing two different cases,
from \eqref{def190} we have 
\begin{equation}
\label{exp100}
\mathbb{P}_i[\rexit]
=
\left\{
\begin{array}{ll}
{\displaystyle
\frac{1}{Z} 
A^{2-L}
\frac{1}{\bar{A}}
\frac{A^i-1}{A-1}
\vphantom{\bigg\{_\big\}}
}
&
\;\;\;
i\le d\\
{\displaystyle
\frac{1}{Z}
A^{2-L}
\Big[
     \frac{1}{\bar{A}}
     \frac{A^d-1}{A-1}
     +
     \frac{1}{A^2}
     \frac{A^{i+1}-A^{d+1}}{A-1}
\Big]
}
&
\;\;\;
i> d.\\
\end{array}
\right.
\end{equation}

Using \eqref{def030}, \eqref{exp080}, and \eqref{exp100} it is possible 
to give an explicit formula for the 
local residence time profile
$\mathbb{E}_i[n_i|\rexit]$ for $i=1,\dots,L-1$.
Finally, using \eqref{def020} we compute the residence time
that we report here without adopting the shorthand notations 
$A$ and $\bar{A}$. We have 
\begin{equation}
\label{exp110}
\begin{array}{rcl}
\Gamma
&\!\!=&\!\!
{\displaystyle
\Big[
     (p-q)^2\Big(-pq+pq\Big(\frac{q}{p}\Big)^L-q\epsilon
     -p\Big(\frac{q}{p}\Big)^L\epsilon
     +\Big(\frac{q}{p}\Big)^d\epsilon\Big)
\Big]^{-1}
\Big(\frac{q}{p}\Big)^{-d}
\phantom{\bigg\{_\big\}}
}
\\
&&\!\!
{\displaystyle
\times
\Big[
     -2pq\Big(\frac{q}{p}\Big)^L\epsilon
     -\Big(\frac{q}{p}\Big)^{2d}
      \Big(
           (1+2d-L)p^2+4pq+(1-2d+L)q^2
      \Big)
      \epsilon
\phantom{\bigg\{_\big\}}
}
\\
&&\!\!
{\displaystyle
\phantom{\times\Big[}
+ 
\Big(\frac{q}{p}\Big)^{d}
\Big(
     pq(p-Lp+q+Lq)+q(3p-Lp+q+Lq)\epsilon
}
\\
&&\!\!
{\displaystyle
\phantom{\times\Big[+\Big(\frac{q}{p}\Big)^{d}\Big(}
     -p
     \Big(\frac{q}{p}\Big)^{L}
     \big((1+L)p(q-\epsilon)
      +q(q-Lq+(L-3)\epsilon)\big)
\Big)
\Big]
.
}
\\
&&\!\!
{\displaystyle
}
\\
\end{array}
\end{equation}
The leading order in $L$, uniformly in the choice of the 
other parameters $d$ and $\epsilon$, is $L/|p-q|$. Thus, the presence 
of the defect does not affect the ballistic
character of the drifted conditioned walk.
It is interesting to remark that this behavior, due to the right 
conditioning, is conserved also for $q>p$, where in a not conditioned 
walk we would expect an exponential behavior with $L$ of the 
mean first hitting time to $L$.

\begin{figure}[!h]
	\centering
	\subfigure{\includegraphics[width=0.475\textwidth]{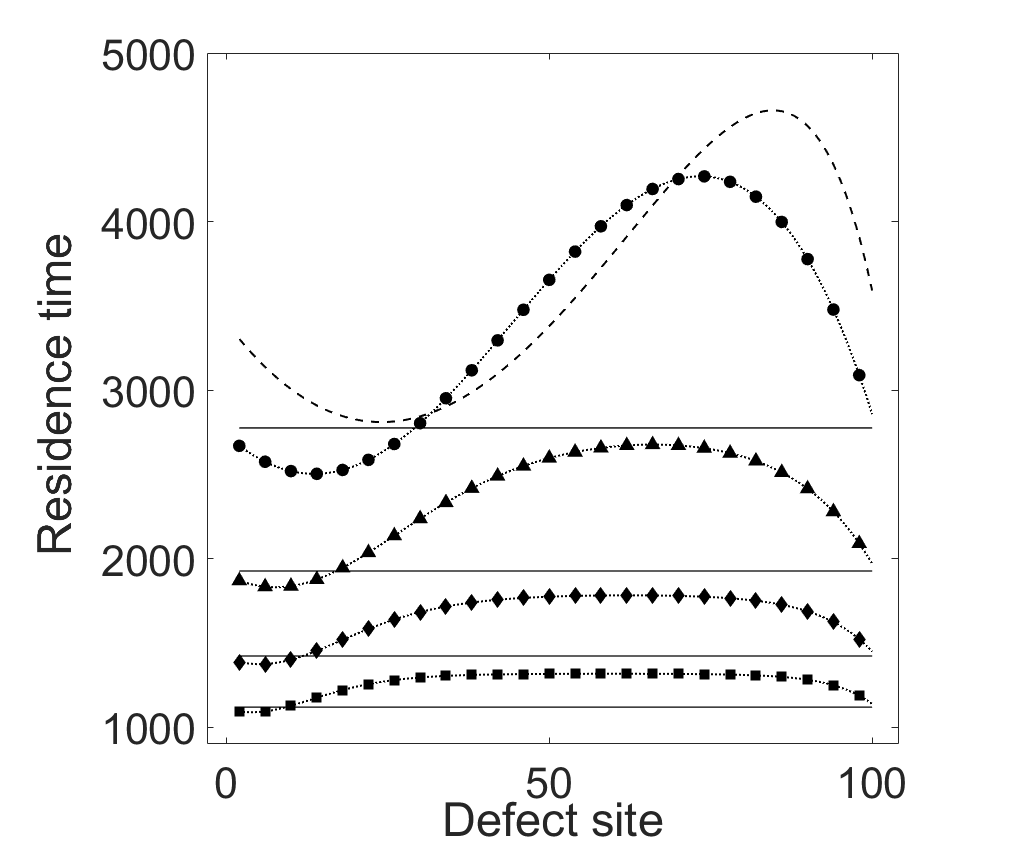}}
	\subfigure{\includegraphics[width=0.475\textwidth]{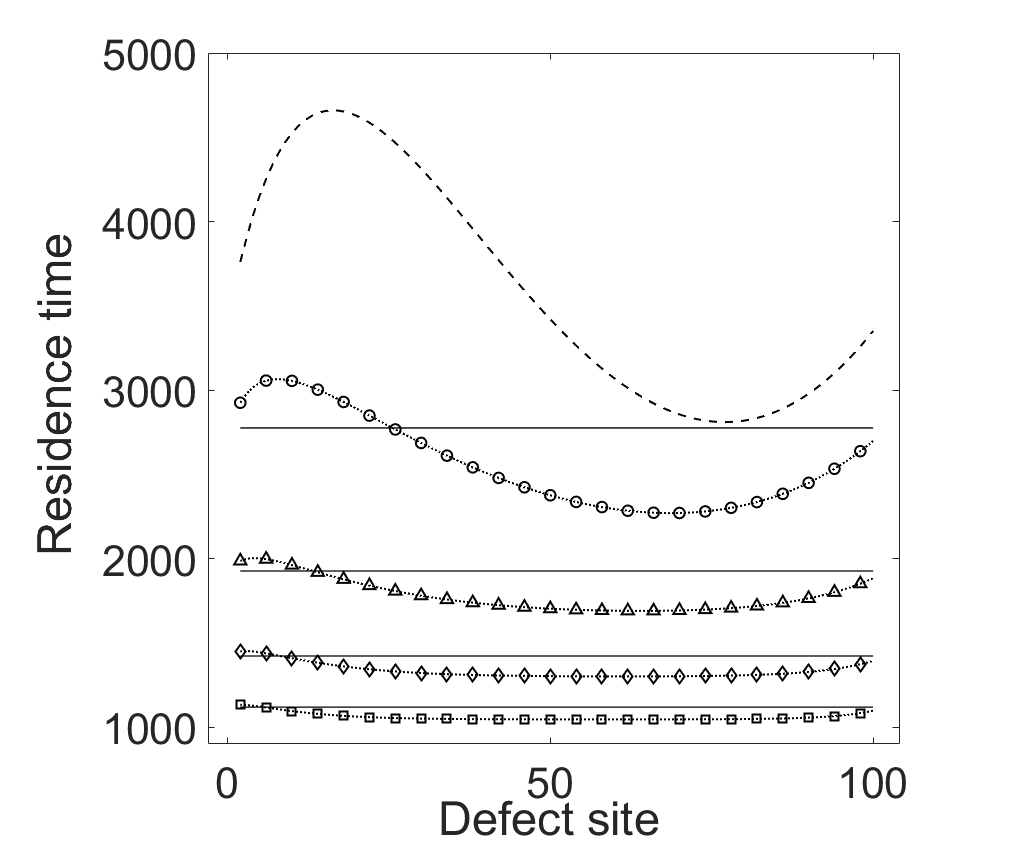}}
	\caption{Residence time as a function of $d$ and of the drift $p-q$, with $p>q$, for a uniformly driven random walk and for $L=101$. The left and right panels refer to the system with $\epsilon=-0.3$ and $\epsilon=0.3$ respectively. In both panels, circles, triangles, diamonds and square refer to $p=0.51,\: 0.52, \: 0.53, \: 0.54$ respectively. Dashed lines refers to the reference residence time for the symmetric random walk (i.e., $p=0.5$), while horizontal solid lines correspond to the residence times of the driven random walks with same values of $p$, but without defects. From the top plot to the bottom one, the no-defect residence times are $\Gamma=2775,\:1926,\:1422,\:1119$ time steps for $p=0.51,\:0.52,\:0.53,\:0.54$ respectively.}
	\label{fig4.2:NullModel_Gamma_driven}
\end{figure}

Fig. \ref{fig4.2:NullModel_Gamma_driven} shows the behavior of the residence time $\Gamma$ in a driven random walk as a function of the defect site $d$. In particular, two cases have been distinguished, both with positive drift ($p-q>0$): the first with $\epsilon<0$ and the second with $\epsilon>0$.

\FloatBarrier
Fig. \ref{fig4.2:NullModel_Gamma_driven} shows how the system with driven random walk and positive drift (i.e., where $p>q$) behaves differently depending on the sign of $\epsilon$, especially for small values of the drift. In particular, in the system characterized by $p>q$ and $\epsilon<0$ it can be noted that for small values of the drift there are intervals of the defect position where the residence time is larger than in the symmetric walk case, despite the drive towards the right exit of the lattice (see Fig.\ref{fig4.2:NullModel_Gamma_driven}, left panel). In both cases ($\epsilon>0$ and $\epsilon<0$), an increasing value of the drift yields a weaker dependence of the residence time on the position of the defect.
In all the cases reported in Fig. \ref{fig4.2:NullModel_Gamma_driven}, the residence times have been computed using formula \eqref{exp110}.

\section{Models with one stochastic defect}
\label{s:move}
In the present work, we are interested in studying the effect of different dynamics of the defect on the mean number of visits and on the local and total residence times. Thus the model with static and fixed defect is modified yielding four different versions. The first two are obtained by assuming that the defect is fixed in space, but it is active ($\epsilon\neq0$) or not ($\epsilon=0$) according to a certain stochastic rule (Models A and B). In the other two, the position of the defect is chosen according to a certain probability distribution over the lattice sites (Models C and D). All the models will be described more in detail and the analytical results obtained in the previous sections are used to derive the mean number of visits and the local and total residence times. These models will be recast as a random walk without defects or with a single static defect through a suitable choice of parameters.

Model A: the defect placed at site $d$ is kept fixed during the random walk, but it is active with probability $\psi$ at each time step. It is immediate to prove that this model generates a behavior which is equivalent to the model with a single static defect at site $d$, where the defect site is characterized by modified jump probabilities, namely $\bar{p}=p+\psi\epsilon$ and $\bar{q}=q-\psi\epsilon$.

Model B: the defect placed at site $d$ is kept fixed during the random walk, but it is active for a random number of time steps (defect is attached to the lattice), sampled from a Poisson distribution of parameter $\lambda_A$ and non-active for a random number of time steps (defect is detached from the lattice), sampled from a Poisson distribution of parameter $\lambda_D$ \cite{MRVMSpre2016}. This model generates an average behavior which is equivalent to that of the fixed and static defect model, where the defect site is characterized by jump probabilities $\bar{p}=p+\frac{\lambda_A}{\lambda_A+\lambda_D}\epsilon$ and $\bar{q}=q-\frac{\lambda_A}{\lambda_A+\lambda_D}\epsilon$.

Model C: the defect position is sampled uniformly over the lattice sites at each time step. This model generates a behavior which is equivalent to that of a driven random walk with jump probabilities $\bar{p}=p+\frac{\epsilon}{L}$ and $\bar{q}=q-\frac{\epsilon}{L}$, evolving through a lattice with no defects. 

Model D: the defect position is sampled at random from a discrete triangular distribution over the lattice sites at each time step. The triangular distribution is characterized by its mode $d$ and its support $[d-a,d+a]$, with $d-a\geq2$ and $d+a\leq L-2$. This model generates a behavior which is equivalent to that of a non-homogeneous random walk with jump probabilities $\bar{p}(i)=p+\beta_i\epsilon$ and $\bar{q}(i)=q-\beta_i\epsilon$, where $\beta_i$ is the probability to find the defect at site $i$.

For Models A and B we shall use results provided in Section \ref{s:sym} if the walk is symmetric and those of Section \ref{s:dri} if the walk is driven. For Model C we use results of Section \ref{s:dri} particularized to the case of no defect. 
Finally, for Model D we will use the general results of Section \ref{s:mod}.

\begin{figure}[!h]
	\centering
	\subfigure{\includegraphics[width=0.475\textwidth]{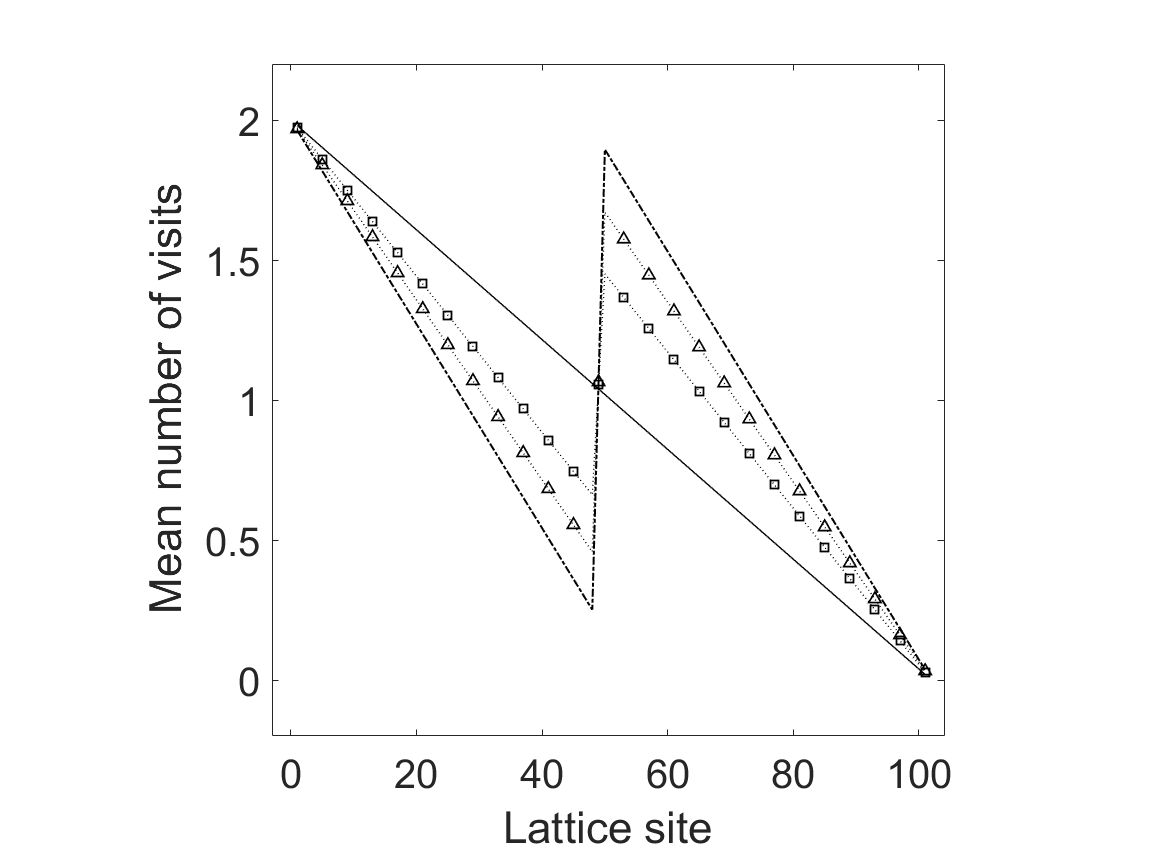}}
	\subfigure{\includegraphics[width=0.475\textwidth]{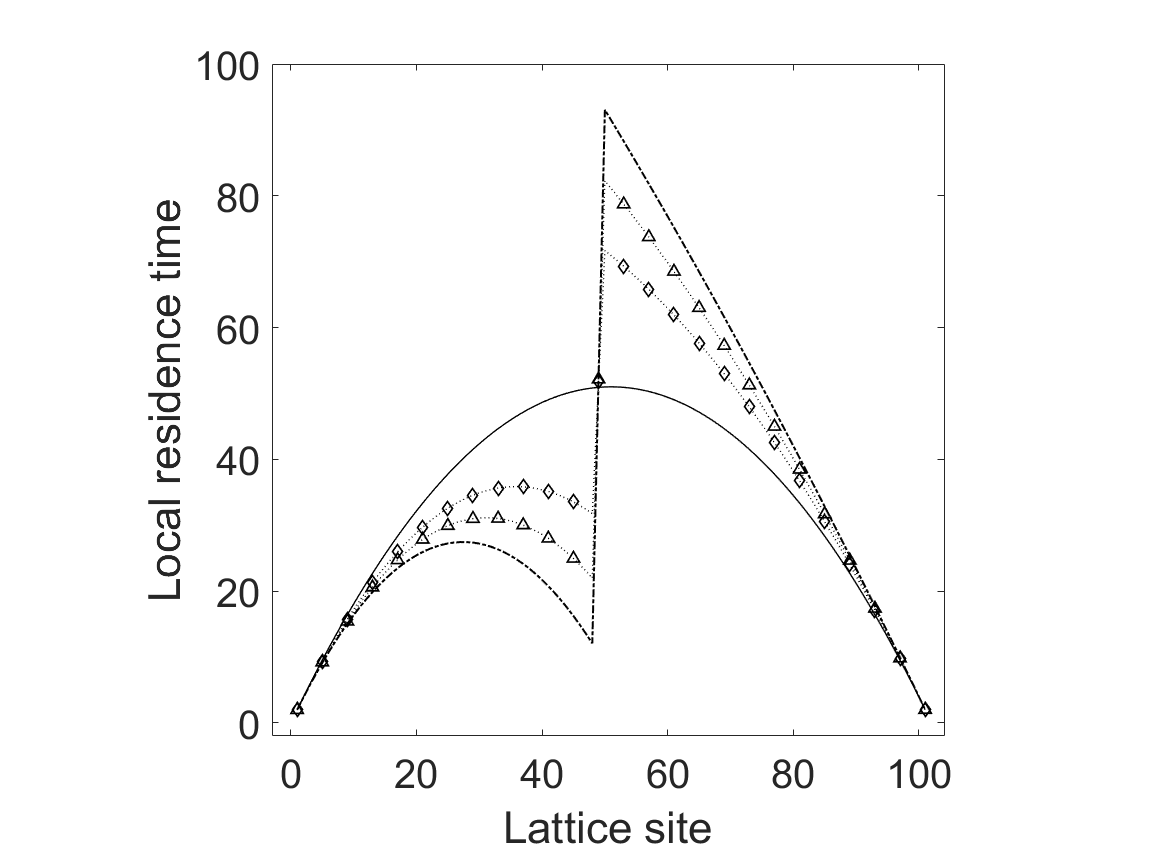}}
	\caption{Mean number of visits and local residence time for Models A and B for $d=51$. Solid lines refer to the random walk model without defects, dash-dotted lines refer to the model with fixed and static defect, empty triangles refer to Model A with $\psi=0.75$, and empty diamonds refer to Model B with $\lambda_A=\lambda_D=100$. In all cases $\epsilon=0.4$ and $L=101$.}
	\label{fig4.3:Dynamics_effect}
\end{figure}

\begin{figure}[!h]
	\centering
	\subfigure{\includegraphics[width=0.475\textwidth]{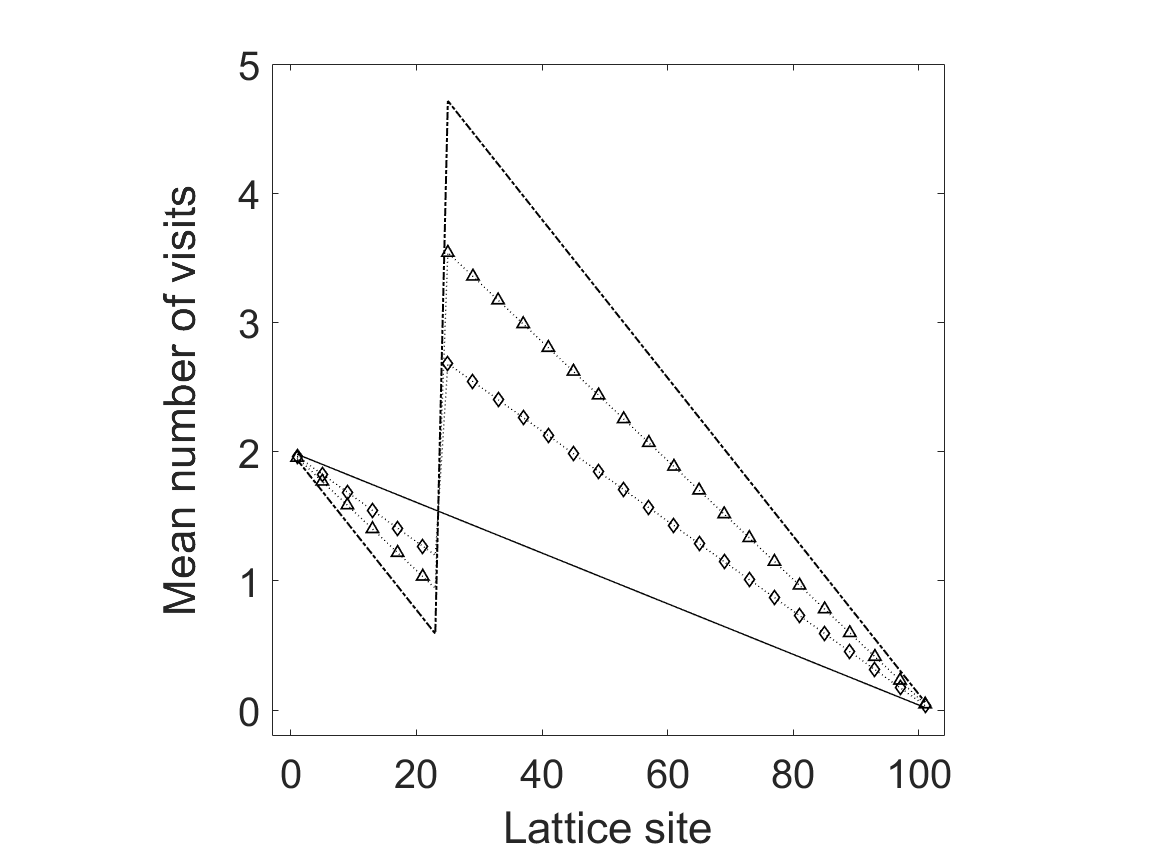}}
	\subfigure{\includegraphics[width=0.475\textwidth]{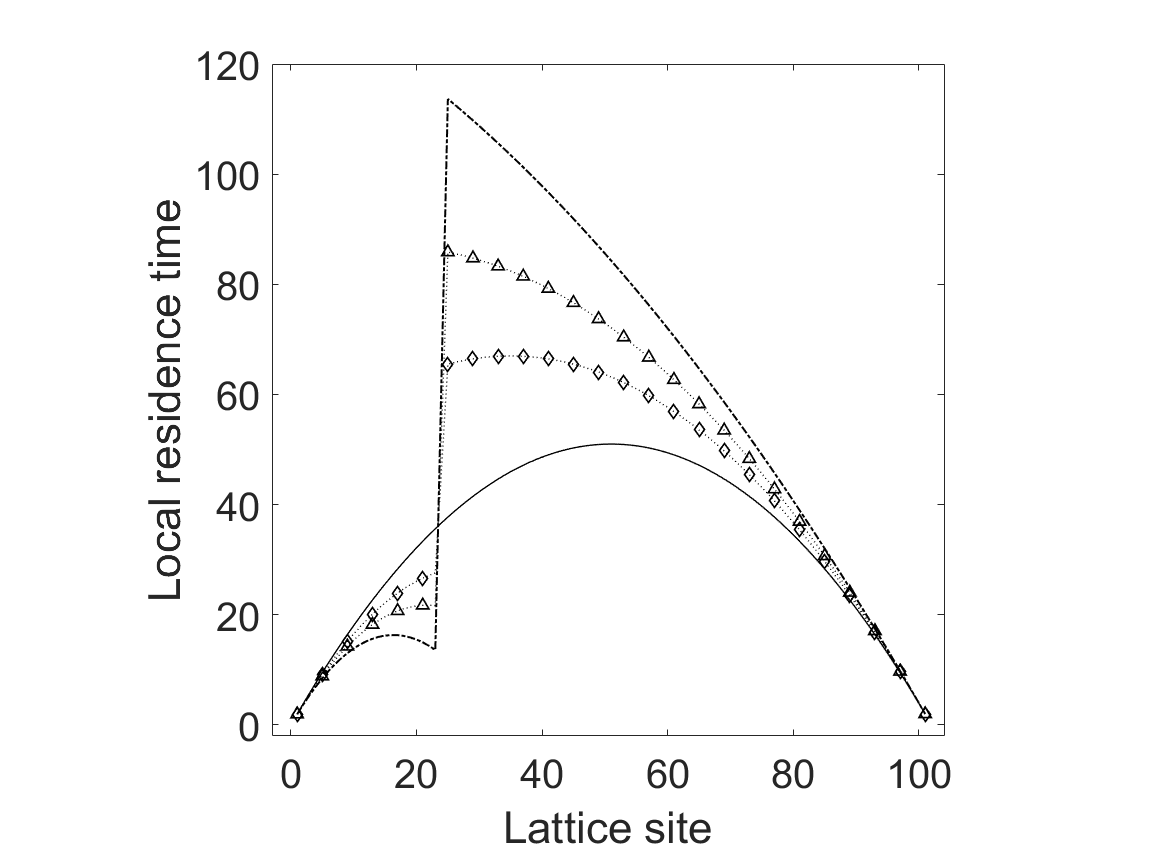}}\\
	\subfigure{\includegraphics[width=0.475\textwidth]{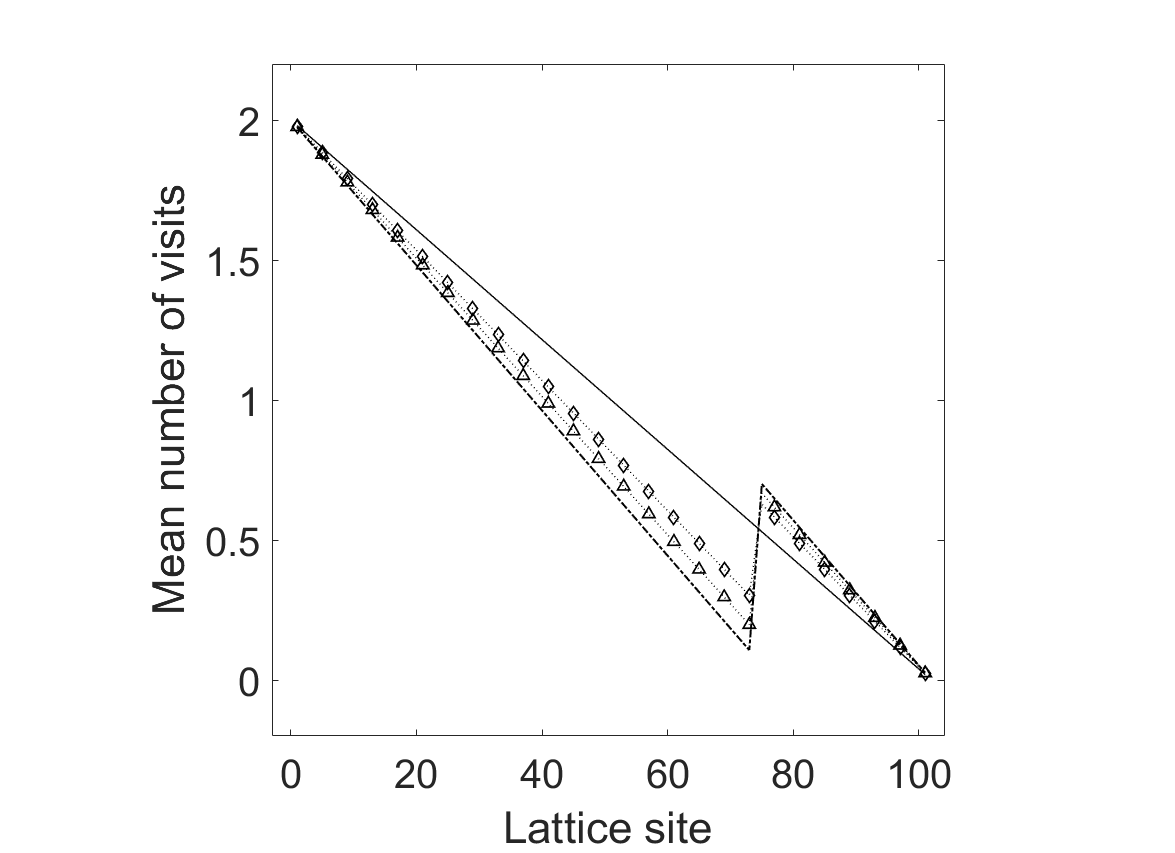}}
	\subfigure{\includegraphics[width=0.475\textwidth]{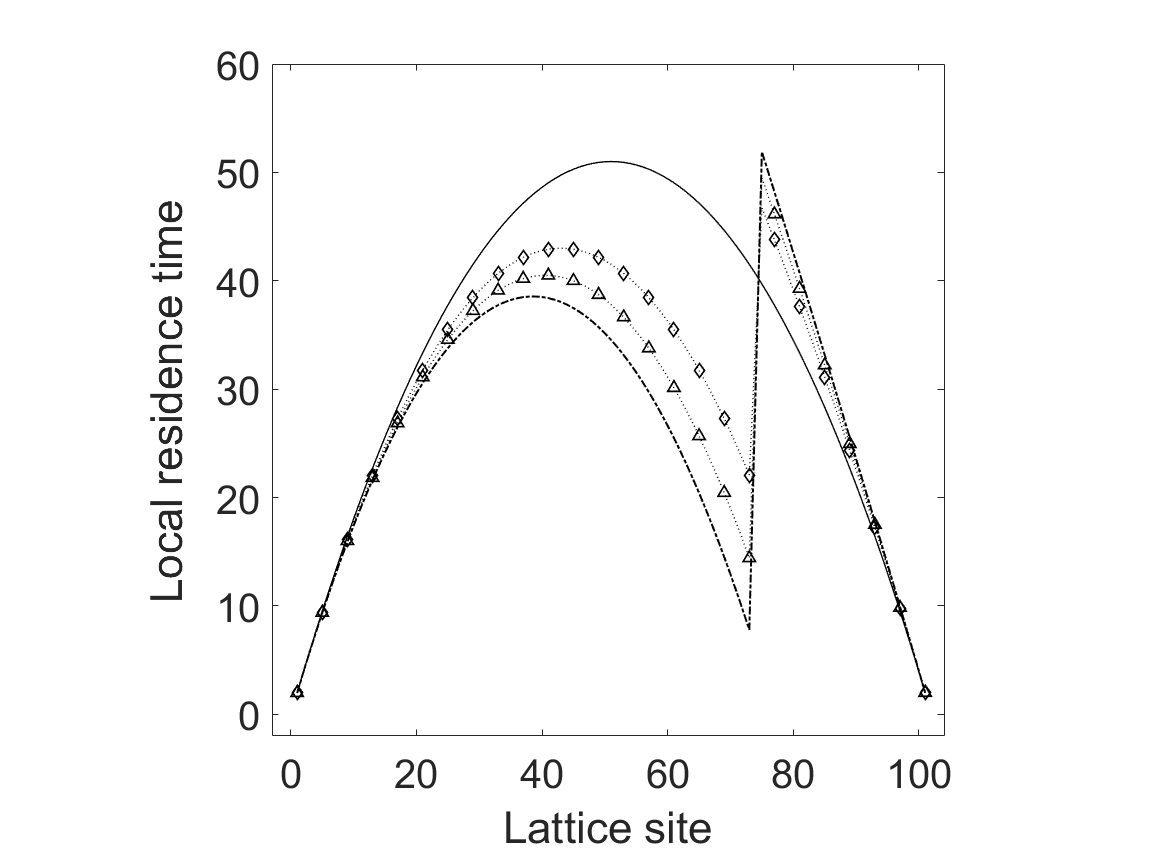}}
	\caption{Mean number of visits and local residence time for Models A and B. From the top to the bottom $d=26$ and $d=76$. Solid lines refer to the random walk model without defects, dash-dotted lines refer to the model with fixed defect, empty triangles refer to Model A with $\psi=0.75$, and empty diamonds refer to Model B with $\lambda_A=\lambda_D=100$. In all cases $\epsilon=0.4$ and $L=101$.}
	\label{fig4.4:Dynamics_effect}
\end{figure}

\FloatBarrier
Fig. \ref{fig4.3:Dynamics_effect} and \ref{fig4.4:Dynamics_effect} show a comparison in terms of mean number of visits and local residence times between the model with fixed and static defect and Models A and B. Details on the parameters are reported in the figures captions. Notice that the mean number of visits and local residence times have been computed using formulae \eqref{exp010} and \eqref{exp050} respectively. The lattice has a odd length to account for the possibility of properly defining a central site. It can be observed that adding a stochastic dynamics to the defect results in a reduction of the amplitude of the jump discontinuity in the mean number of visits and in the local residence times profiles. Another remarkable effect of the defect is that on the local residence times (see Fig. \ref{fig4.3:Dynamics_effect}, right panels). In fact, the values of the local residence times and the amplitude of the jump discontinuity are remarkably higher when $d=26$ than when $d=76$. Recalling equation \ref{def020}, it can be seen that this behavior is in agreement to that of $\Gamma$ as a function of $d$ in Fig. \ref{fig4.1:NullModel_Gamma_NONdriven} when $\epsilon>0$, where the asymmetric effect of the defect position on the total residence time is stronger on the left half of the lattice than it is on the right half. In general, residence times for Model A and B behave exactly as those in the fixed and static defect model, except for a proper scaling of $\Gamma$ due to the reduced effectiveness of the defect site (whose intensity is diminished by a factor of $\psi$ or $\frac{\lambda_A}{\lambda_A+\lambda_D}$ for Model A and B respectively).

\begin{figure}[!h]
	\centering
	\subfigure{\includegraphics[width=0.475\textwidth]{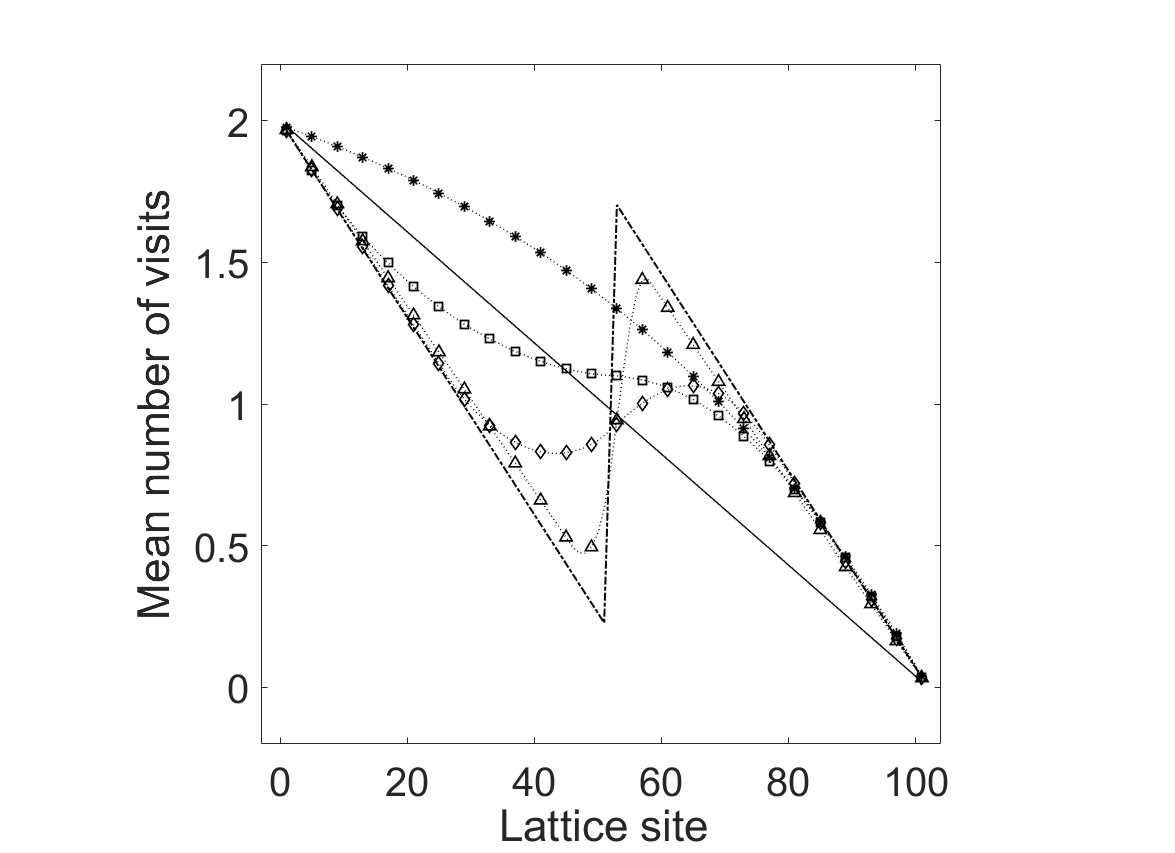}}
	\subfigure{\includegraphics[width=0.475\textwidth]{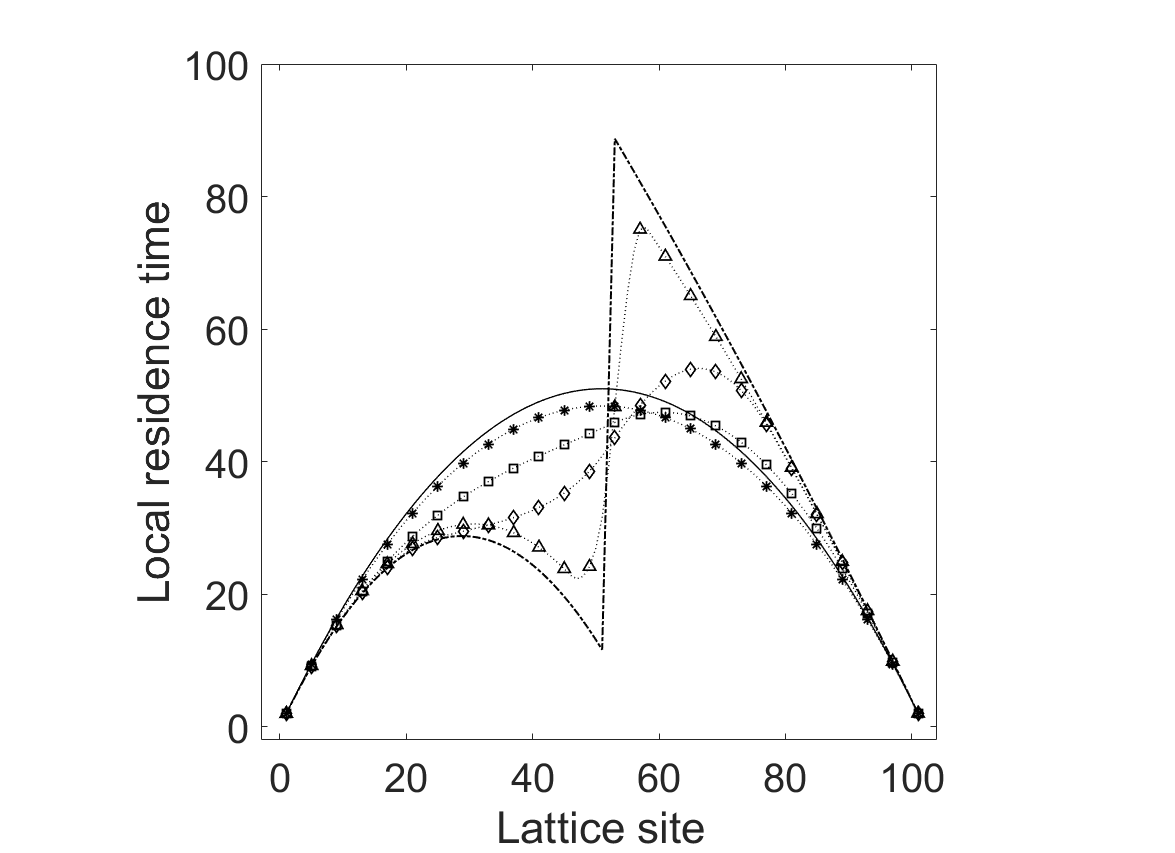}}	
	\caption{Mean number of visits and local residence times for Model D for different values of $a$ and same metrics for Model C. Dash-dotted lines and asterisks refer to the model with fixed and static defect and to Model C respectively. Triangles, diamonds and squares refer to Model D with $a=5,\: 25,\: 49$ respectively. All the models are characterized by $\epsilon=0.4$, $d=51$ and $L=101$. Solid lines refer to the no-defect model. }
	\label{fig4.5:Motion_effect}
\end{figure}
\FloatBarrier

In Fig. \ref{fig4.5:Motion_effect} the effect of the motion of the defect in terms of mean number of visits and local residence times is reported, comparing the model with fixed and static defect with Models C and D. The mean number of visits and local residence times corresponding to the fixed and static defect model and to Model C have been computed through formulae \eqref{exp010}, \eqref{exp050}, \eqref{exp080} and \eqref{exp100} respectively. The profiles for Model D have been obtained through the general formulae \eqref{def130} and \eqref{def030} where \eqref{def190} has been used for the right-exit probability.  

In this case, the main effect of the defect motion on the mean number of visits and on the local residence times is to smooth out the profiles and produce a transition from the discontinuous piecewise-linear behavior of the model with fixed defect to the continuous, smooth profile related to Model C, where the defect can be found uniformly at random in every site of the lattice. In particular, taken Model D, the case $a=0$ ideally coincides with the model with fixed defect, since the defect has probability $1$ to be placed in a specific site.

\begin{figure}[!h]
	\centering
	\subfigure{\includegraphics[width=0.325\textwidth]{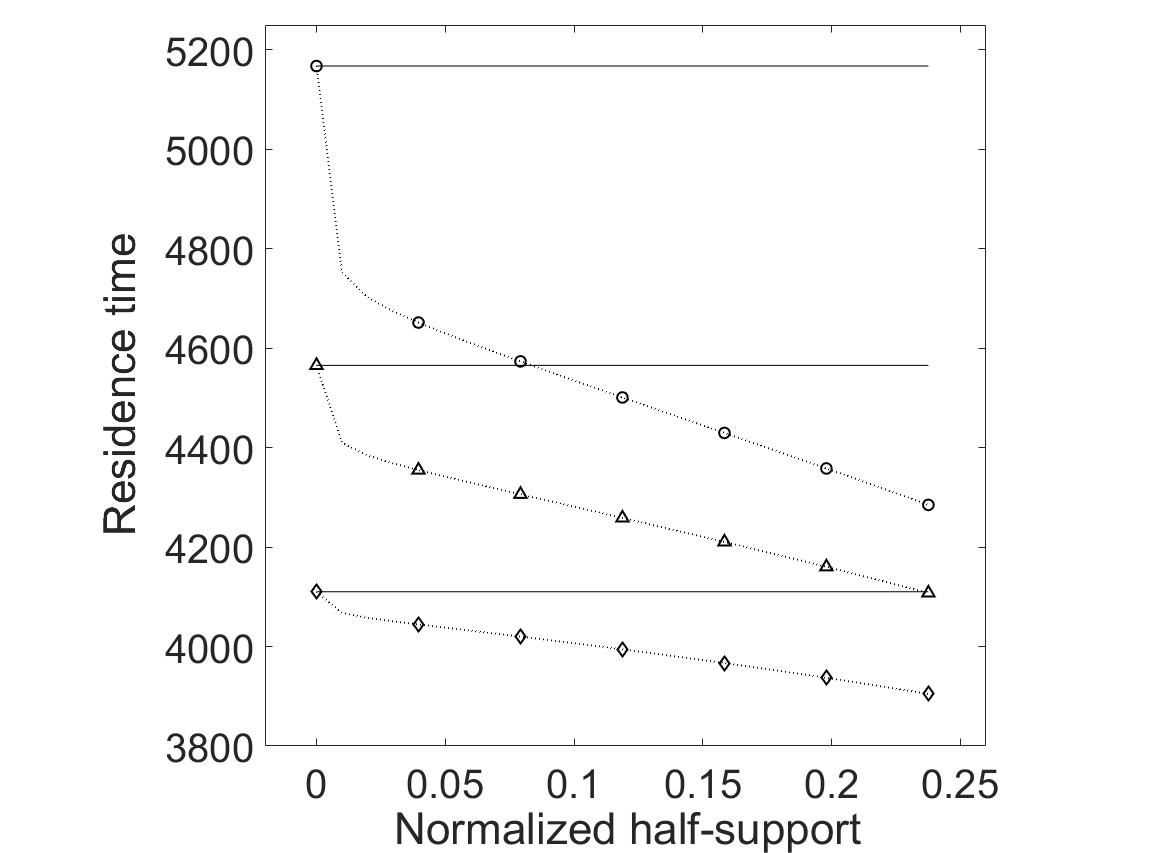}}
    \subfigure{\includegraphics[width=0.3175\textwidth]{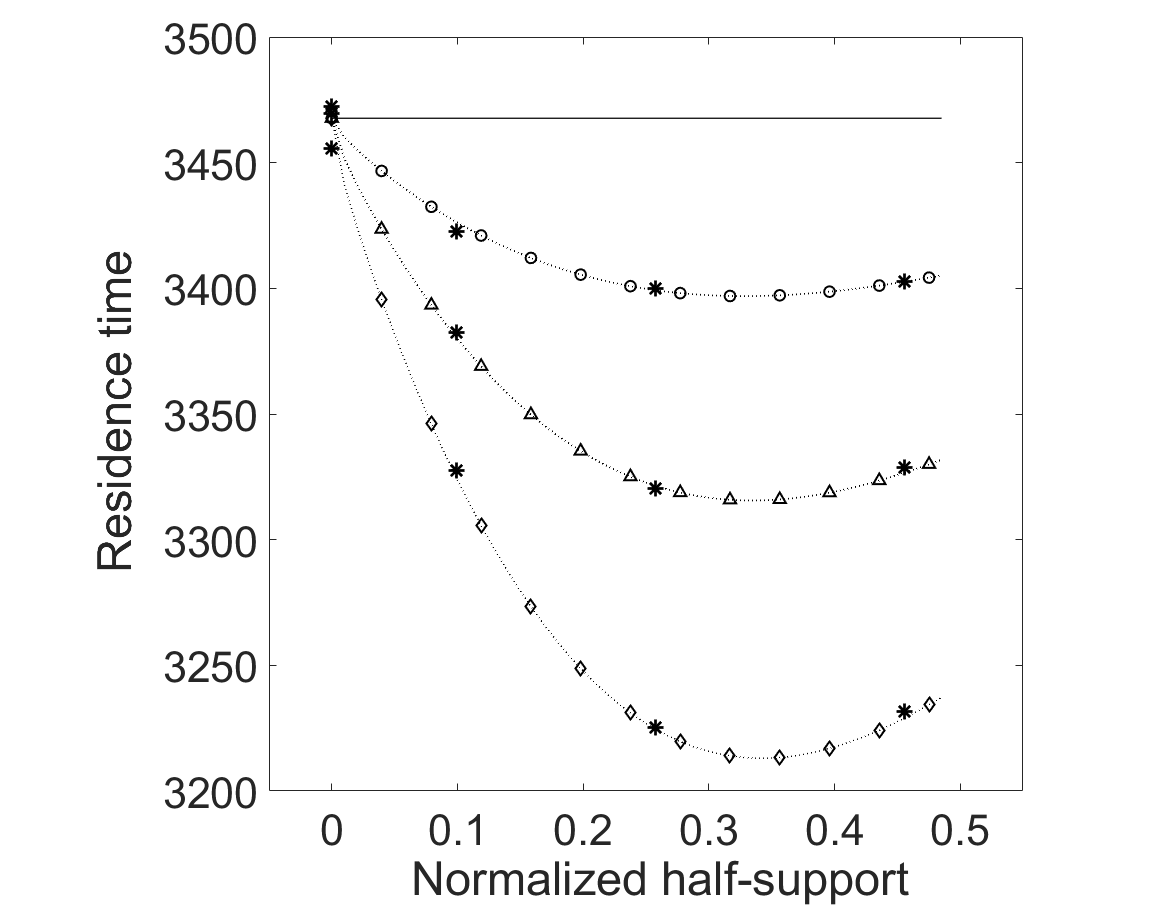}}
	\subfigure{\includegraphics[width=0.325\textwidth]{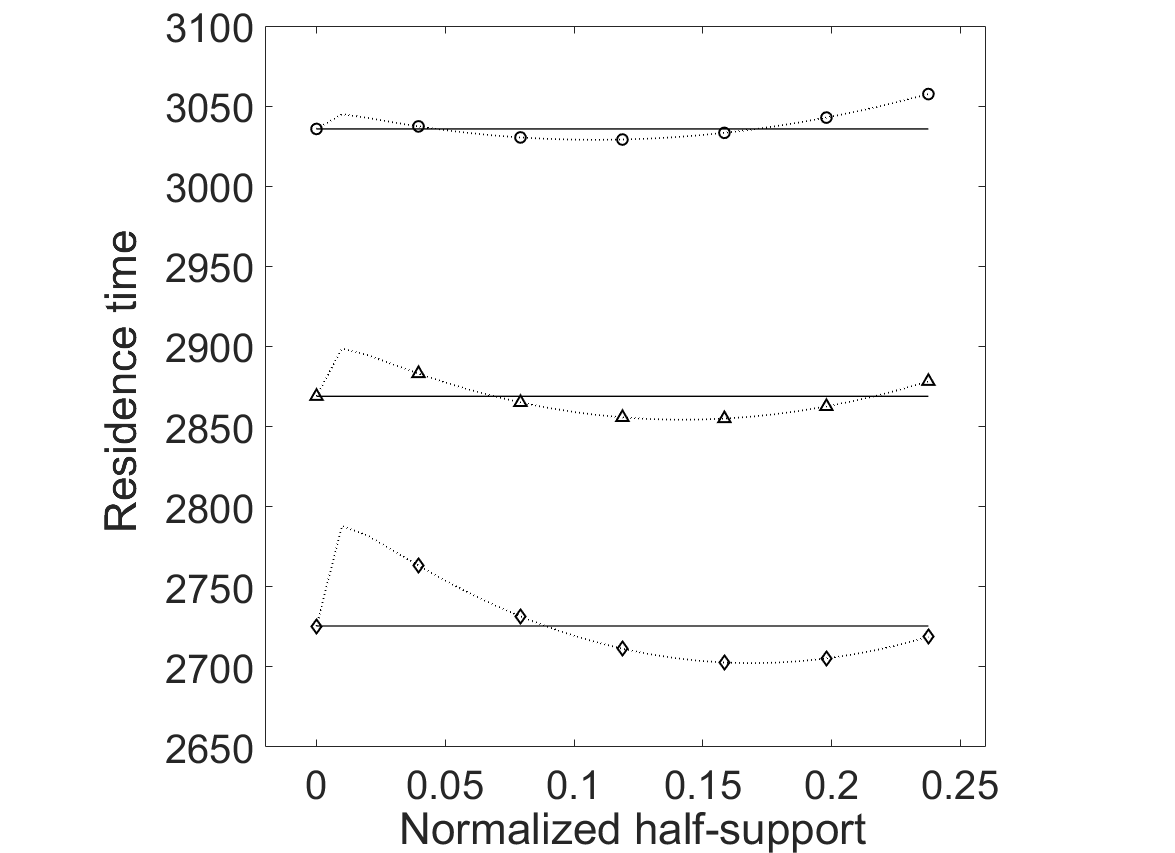}}
	\caption{Residence times for Model D as functions of $a/L$. From the left to the right, Model D with mode $d=25,\: 51,\: 75$. In all the panels $L=101$ and empty circles, triangles and diamonds refer to defect intensities of $\epsilon=0.2,\: 0.3,\: 0.4$ respectively. The solid horizontal lines correspond to residence times in the model with fixed and static defect, namely $\Gamma=5167.5,\: 4565.4,\: 4110.3$ time steps for $d=25$ and $\epsilon=0.2,\: 0.3,\: 0.4$ respectively, $\Gamma=3467.7$ time steps for $d=51$ and $\Gamma=3035.8, \: 2868.9, \: 2725.5$ time steps for $d=75$ and $\epsilon=0.2,\: 0.3,\: 0.4$ respectively. Asterisks in central panel indicate the residence times obtained from simulations.}
	\label{fig4.6:Movement_effect_Gamma}
\end{figure}
\FloatBarrier

A noticeable aspect is the effect of the defect motion on the residence time $\Gamma$. It can be observed that, independently of the value of $\epsilon$ (with $\epsilon>0$) and from the width of the triangular distribution (parameterized by the width normalized to the lattice length, namely $a/L$), the residence time in presence of a moving defect is always shorter than that of the system with fixed defect, at least when the center $d$ of the distribution coincides with the central lattice site or it is placed at its left, see left and central panels in Fig. \ref{fig4.6:Movement_effect_Gamma}. When $\epsilon>0$ and the center of the distribution is placed in the right half of the lattice, there is only a finite interval of values of $a/L$ for which the residence time in presence of the moving defect is shorter than the corresponding for the fixed defect, as can be seen from the right panel in Fig. \ref{fig4.6:Movement_effect_Gamma}. 

The most important observation about this model is the residence time invariance with respect to the sign of $\epsilon$ when $\Gamma$ is a function of the defect motion (to be intended as the triangular distribution spread $a$). In fact, the behavior of $\Gamma$ shown in Fig. \ref{fig4.6:Movement_effect_Gamma} is also obtained for $\epsilon<0$, except for the fact that the behavior observed for $d=25$ in the case $\epsilon>0$ is recovered for $d=75$ and $\epsilon<0$ and vice-versa. This symmetry with respect to the central site of the lattice for different signs of $\epsilon$ is of the same nature of that shown in Fig. \ref{fig4.1:NullModel_Gamma_NONdriven} and can be traced back to the effects of the hopping probabilities $p_k$ and $q_k$ on formulae \eqref{def130}, \eqref{def140} and \eqref{def190}. More specifically, this result can be explained as follows: defect moving around the central lattice site and having positive intensity will increase the local residence time in the sites to the right of the center and decrease it in those to the left; on the other hand, defect moving around the central lattice site and with negative intensity will have the opposite effect. In these two systems, the local residence times will be distributed differently (in particular, they will be symmetric with respect to the lattice center) but they will sum up to the same value of $\Gamma$.

These remarkable results on the behavior of such a system have been tested against computer simulations (see Fig. \ref{fig4.6:Movement_effect_Gamma}, central panel, for the residence times resulting from simulations). In particular, a random walk on the line algorithm has been implemented which takes into account the dynamics of the defect site. A large enough number ($15\cdot10^6$) of walks have been simulated and averages over the resulting set of trajectories have been computed. As shown in the central panel of Fig. \ref{fig4.6:Movement_effect_Gamma}, an optimal agreement between simulations and theory is obtained for all the tested values of $\epsilon$ and $a$. 






\section{Discussion and conclusions}

\label{s:con}


We have studied the residence time in the framework of a simple random walk
on a 1D lattice that we called lane. 
By using the theory of absorbing Markov chains we have first derived
the expression \eqref{def040}, in which the residence time 
is written in terms of the probability that the walker started at a 
generic site $i$ exits the lane through the right exit at $L$ and 
the mean number of visits that the walker pays at site $i$. 
For these building bricks we have found general expressions,
in Sections~\ref{s:vis} and \ref{s:rep}, which are valid for any choice 
of the left an right jump probabilities, $q_k$ and $p_k$, defining 
the walk.

These results have then been used in Section~\ref{s:explicit} to 
write the residence time in the case of a homogeneous walk with a 
single static defect. In particular, in Sections~\ref{s:sym} and 
\ref{s:dri}, we have derived explicit expressions for the residence 
time in the symmetric and in the driven case. Those expressions 
revealed to be very useful in the following sections to study cases 
in which a defect dynamic is considered. In particular we have 
shown that adding a dynamic to the defects induces a smoothing of the 
observable behaviors that present abrupt jumps in the case 
of a static defect.

Although the simple random walk is a very basic model, 
we have seen how it can provide general interesting insights 
for the residence time behavior, such as the large $L$ behavior.
One of the interesting features of studying simplified models is 
the fact that they can help to shed some light on the 
behavior of much more complex systems. We want to discuss two 
interesting cases coming from two very different applicative contexts.

We first consider the crowd dynamics problem introduced in 
the paper \cite{WG2021}: pedestrians move from the left to the right 
through a rectangular corridor avoiding other pedestrians which 
form vertical queues, namely, queues orthogonal with respect to 
the direction of motion of the passing pedestrians.
The model introduced in this paper is very complicated, many 
parameters accounting for the several interaction acting on the 
pedestrians are considered, and it is used to study in detail 
the different effects that enter in the computation of the residence 
time, i.e., the mean time needed by the passing pedestrian to cross the 
corridor. In particular the authors compute the quantity $I_\textrm{pas}$, 
which is 
defined as the ratio between the residence time in absence of 
queues and the residence time in presence of queues.
In particular, in \cite[Fig.~8]{WG2021}, the authors study 
its behavior as a function of the parameter $\phi$, which is 
the parameter 
weighting the repulsive 
force that the passing pedestrians exert on the queuing ones.

In our simplified modelling, the walker can be thought as the 
passing pedestrian and the effect of the queuing pedestrians 
on it can be reduced to the presence of the defect site. 
Our model can be thought as a sort of effective model for the passing 
pedestrian problem with the parameter $\epsilon$, associated with 
the defect, containing all the informations on the interactions 
conditioning the motion of the passing pedestrians. For instance, 
coming back to \cite[Fig.~8]{WG2021}, it is possible to use our 
explicit formula \eqref{exp110}
for the residence time in the driven case to fit 
the data of \cite[Fig.~8]{WG2021} assuming that the effective 
$\epsilon$ is related to the parameter $\phi$ by the very general 
function

\begin{equation}
\label{co000}
\epsilon(\phi)=\bar{a}\frac{\phi^\alpha}{\phi^\alpha+c}-\bar{b}.
\end{equation}

\begin{figure}[!h]	
	\centering	
	\includegraphics[width=0.475\textwidth]{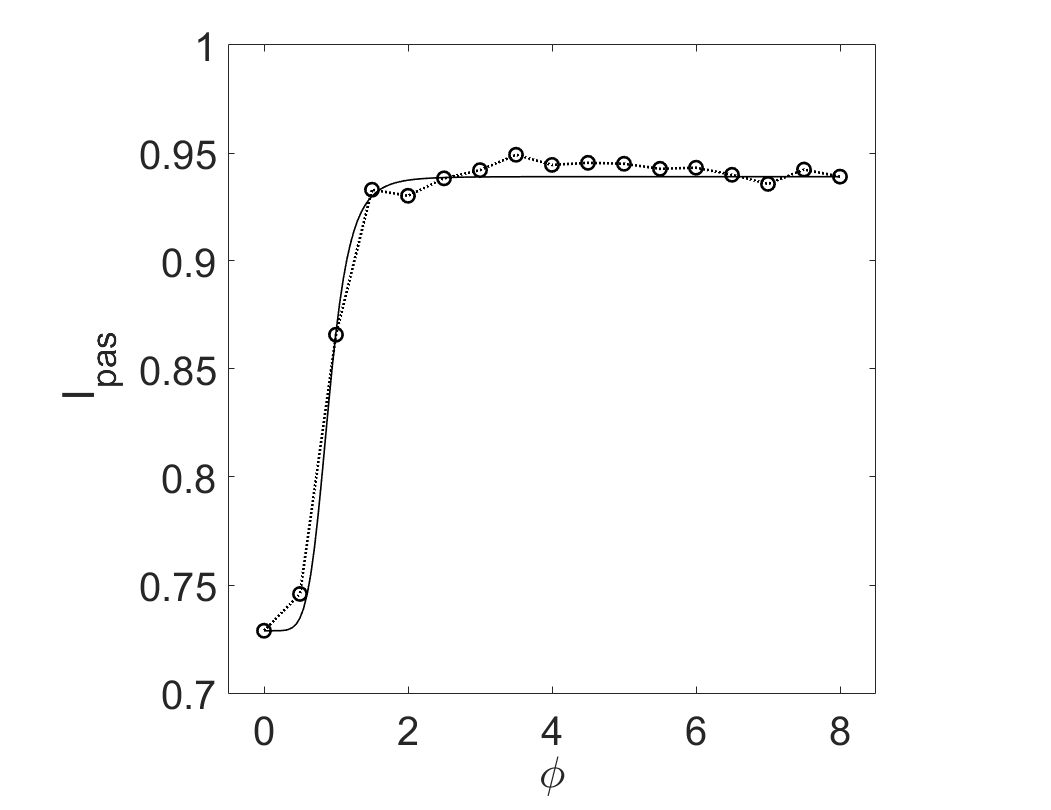}
	\caption{Fitting of the values of the efficiency parameter $I_\textrm{pas} $ as extracted from \cite[Fig.~8]{WG2021}. 
		Dotted line with empty circles represents the data to be fitted, while the solid line is the analytical $I_\textrm{pas}$ as a function of
	$\epsilon(\phi)$ obtained by using formula \eqref{exp110} for the residence times and fitting the parameters in formula \eqref{co000}.}	
	\label{fig6.1:WuetAl_Ipas_fit}	
\end{figure}

If we consider, in our model, $p=0.55$ and $q=0.45$, 
from the extreme data in the picture 
we have that 
$\bar{a}=0.2230$ and $\bar{b}=0.4152$. 
Performing the best fit we find 
$\alpha=6.043$ and $c=1.074$. The comparison between our prediction and 
the data in \cite{WG2021} is depicted in Fig.\ref{fig6.1:WuetAl_Ipas_fit}.

\FloatBarrier
Equation~\eqref{co000} gives the effective value, as a function of 
the parameter $\phi$, of the decrease in probability that the passing 
pedestrian crosses the queue when it reaches it. 
It is worth nothing that this result is robust in the sense that the 
fitting parameters $\alpha$ and $c$ appear to depend poorly 
on the choice of the left and right jumping probabilities $p$ and $q$
in the simple random walk model. 

Another relevant application of the residence time idea is related 
to stormwater runoff ponds design. Water running off from residential and 
urban areas, possibly after severe meteorological events like storms 
and heavy rainfalls, is known to drain and transport lots of different 
pollutants, which may result in a serious threat for water or marine 
ecosystems. To prevent drained pollutants to reach delicate water 
ecosystems, natural and artificial ponds are used to accelerate the 
deposition and absorption of undesired substances and components. 
As it has been demonstrated \cite{walker1998modelling}, residence times 
of the water flow inside the ponds are of crucial impact on their 
effectiveness. In particular, the higher the residence time, the 
higher is the pond efficiency in terms of deposition of 
pollutants \cite{Hol04}. Different design strategies, 
aimed at increasing the efficacy of the ponds, have been proposed. 
For instance, the installation of baffles, underwater berms or surface 
islands, acting as obstacles to the water flow, can increase significantly 
the residence time \cite{khan2009modeling,agunwamba1992field}.

Design aspects, like the maximization of the residence time, 
are studied in the literature by means of fluidodynamic approaches
\cite{khan2009modeling}. It is very interesting to remark that
the qualitative behavior of those results can be interpreted 
by using our random walk model. 
We analyze, for instance, the results discussed 
in \cite{khan2009modeling} where 
a rectangular pond is considered and the water inlet and outlet are 
placed symmetrically on the shortest sides of the pond. 
In particular we focus on the results reported in 
\cite[Fig.~5]{khan2009modeling}, where 
RTD curves are plotted versus time in five cases: 
pond with no baffle (case 1), 
pond with an island placed at one quarter of pond length from the 
inlet (case 5),
pond with an island placed at one tenth of pond length from the 
inlet (case 6), pond with a subsurface bern placed at one quarter of pond length 
from the inlet (case 7),
pond with a subsurface bern placed at half pond length 
from the inlet (case 8).
Those curves have been found by solving numerically suitable 
fluidodynamic equations.

These geometries are compatible with our 1D model, since 
the islands and the berns act symmetrically along the direction 
parallel to the shortest sides of the pond, namely, the direction orthogonal 
to the main water motion.

Data in \cite[Fig.~5]{khan2009modeling} show
that the residence time in case 5 is larger than that in case 6, which, 
on turn, is larger than the residence time in case 1. This is precisely 
what we observe in our random walk driven model discussed in Section~\ref{s:dri}. 
Indeed, if one focuses on any of the curves with larger drift 
plotted in the left panel of Fig.~\ref{fig4.2:NullModel_Gamma_driven}, 
one can see that the residence time increases when a defect is placed 
at the site $10$ (one tenth of the length of the lane) and 
increases even more when it is placed at the site $25$ (one fourth of the 
length of the lane). 

\begin{figure}[!h]	
	\centering	
	\includegraphics[width=0.475\textwidth]{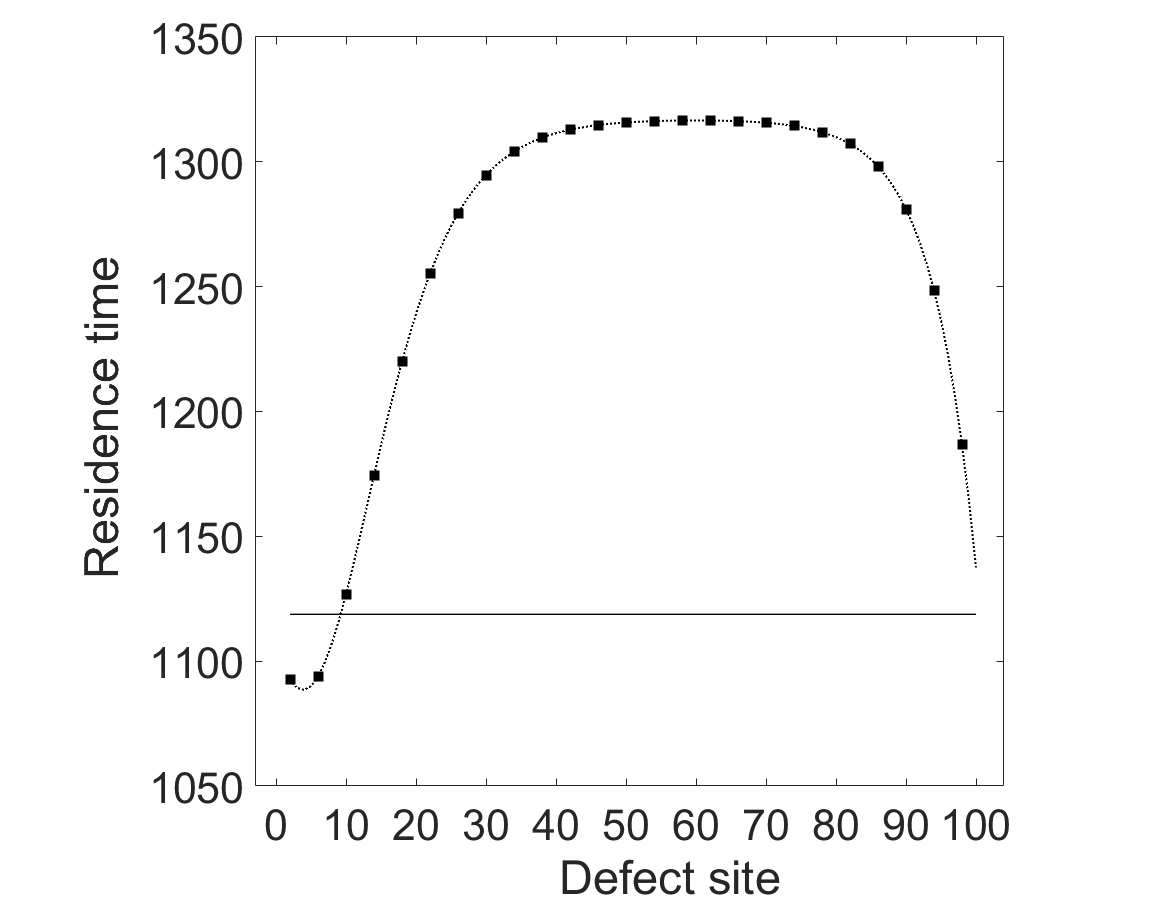}
	\caption{Residence time as a function of $d$ for a uniformly driven random walk with $p=0.54$, $\epsilon=-0.3$ and $L=101$. Dotted line with plain squares refers to the residence time of the walker for varying defect site. Horizontal solid line corresponds to the residence time of the driven random walk with same value of $p$, but without defects. The no-defect residence time is $\Gamma=1119$ time steps, while for $d=10$ and $d=25$ the residence times are $\Gamma=1127,\:1274$ time steps respectively.}	
	\label{fig:DrivenRW_runoffponds}	
\end{figure}

In particular, Fig. \ref{fig:DrivenRW_runoffponds} shows the case in which $p=0.54$,
where this behavior is more visible.

The random walk model does not appear to be completely 
effective when berns are considered.
Indeed, data in \cite[Fig.~5]{khan2009modeling} show
that the residence time in case 7 (very similar to the one measured in 
case 5) is larger than that in case 8, which, 
on turn, is larger than the residence time in case 1. The 
fact that the residence time in case 7 is larger than that in 
case 1 is compatible with our results 
plotted in the left panel of Fig.~\ref{fig4.2:NullModel_Gamma_driven}. 
On the other hand, based on our analysis, we would have expected that 
the residence time in case 8 would have been very similar to the one 
measured in case 7, but this seems not to be the case.

Summarizing, we have developed a rather complete theory of the 
residence time for the random walk in presence of defects. In 
particular, we have found explicit expressions in the case of a 
single defect, both in the symmetric and in the driven case. 
We have discussed our results in the framework of our abstract model 
and we have also shown how they can help, to some extent, to understand the 
behavior of much more complex systems. 


\appendix
\renewcommand{\theequation}{\Alph{section}.\arabic{equation}}

\section{Analogy with the stationary occupation number profile}
\label{a:occup}

\par\noindent
We consider a system of many independent particles performing a 
random walk on $1,\dots, L-1$ in continuous time. We suppose particles 
move with rate one according to the jump probabilities introduced in 
Section~\ref{s:mod}. Moreover,  particles are introduced with 
rate $\alpha$ at site $1$ and $\delta$ at site $L-1$. 

This system can be recast as a Zero Range Process with open 
boundaries and intensity 
function associated with site $i$ equal to the number of 
particles occupying such a site \cite{EH2005,CCpre2017}. 

The mean value of particles occupying each site of the lane $1,\dots,L-1$ 
at stationarity satisfy a set of recursive equation identical 
to equations \eqref{def120} provided we choose $\alpha=1$ and 
$\delta=0$. See, for instance, 
\cite[equation~(13)]{LMS2005},
where the equations are reported for 
a case in which the jump probabilites are spatially homogeneous, 
or 
\cite[equation~(14)]{CCpre2017}, 
where the equations are again reported for 
a case in which the jump probabilites are spatially homogeneous
but the presence of a defect site is taken into account.
The generalization of the formulae in 
\cite{LMS2005,CCpre2017}
to the case of arbitrary jumping 
probabilities is straightforward.

Thus, we can conclude that the mean number of visits profile 
of a random walk with two absorbing site coincides with 
the stationary occupation number profile of the independent particles 
Zero Range Process with same jumping probabilities and inlet rate $1$ 
on the left--end of the lane and $0$ on the right--end of the lane.

In \cite{CCpre2017} the stationary occupation number profile 
has been explicitly computed in the symmetric case discussed here in 
Section~\ref{s:sym} for the defect at the center of the lane. 
Indeed, equations \eqref{exp000} and \eqref{exp010}
reduces to \cite[equations~(20) and (21)]{CCpre2017} 
provided we choose the parameters as follows: $L-1=2R+1$, 
$d=R+1$, $\alpha=1$, and $\delta=0$.

\end{document}